\documentclass[12pt,preprint]{aastex}

\newcommand{\eg}{{\it e.g.}\ }

\newcommand{\ie}{{\it i.e.}\ }

\newcommand{\etal}{{\it et al.}\ }

\newcommand{\ps}{Pan-STARRS}

\newcommand{\psone}{PS1}

\def\deg{\hbox{$^\circ$}}

\begin{document}

\title{Detection of Earth-impacting asteroids \\
with the next generation all-sky surveys}

\author{
Peter Vere\v{s}\altaffilmark{1},
Robert Jedicke\altaffilmark{2},
Richard Wainscoat\altaffilmark{2},
Mikael Granvik\altaffilmark{2},\\
Steve Chesley\altaffilmark{3},
Shinsuke Abe\altaffilmark{4},
Larry Denneau\altaffilmark{2},
Tommy Grav\altaffilmark{5},
}

\altaffiltext{1}{Faculty of Mathematics, Physics and Informatics,
Comenius University, Mlynska Dolina, 842 48 Bratislava, Slovakia}

\altaffiltext{2}{University of Hawaii, Institute for Astronomy, 2680
Woodlawn Drive, Honolulu, HI 96822-1897, USA}

\altaffiltext{3}{Jet Propulsion Laboratory, California Institute of
Technology, Pasadena, CA 91109, USA}

\altaffiltext{4}{Institute of Astronomy, National Central University,
No. 300, Jhongda Rd, Jhongli City, Taoyuan County 320, Taiwan}

\altaffiltext{5}{Department of Physics and Astronomy, Bloomberg 243,
Johns Hopkins University, 3400 N. Charles St., Baltimore, MD,
21218-2686, USA}

\shorttitle{Detection of Earth-impacting asteroids}
\shortauthors{Veres et al.}

\begin{abstract}
We have performed a simulation of a next generation sky survey's
(Pan-STARRS 1) efficiency for detecting Earth-impacting asteroids.
The steady-state sky-plane distribution of the impactors long before
impact is concentrated towards small solar elongations
\citep{Chesley04} but we find that there is interesting and
potentially exploitable behavior in the sky-plane distribution in the
months leading up to impact.  The next generation surveys will find
most of the dangerous impactors ($>$140~m diameter) during their
decade-long survey missions though there is the potential to miss
difficult objects with long synodic periods appearing in the direction
of the Sun, as well as objects with long orbital periods that spend
much of their time far from the Sun and Earth.  A space-based platform
that can observe close to the Sun may be needed to identify many of
the potential impactors that spend much of their time interior to the
Earth's orbit.  The next generation surveys have a good chance of
imaging a bolide like 2008~TC$_3$ before it enters the atmosphere but
the difficulty will lie in obtaining enough images in advance of
impact to allow an accurate pre-impact orbit to be computed.
\end{abstract}

\maketitle

\section{Introduction}

Throughout most of human history it was not understood that the
Earth has been battered by large asteroids and comets and
that the impacts and subsequent environmental changes have serious
consequences for the survival and evolution of life on the planet.
But in the past $\sim$50 years more than 170 impact structures
have been identified on the surface of the Earth
\citep{Impact.Database08}. Were it not for the Earth's protective
atmosphere, oceans, erosion and plate tectonics, the surface of
the Earth would be saturated with impact craters like most other
atmosphereless solid bodies in our solar system.  While the impact
probability is now relatively well understood as a function of the
impactor size \citep[\eg][]{Brown02,Harris07} this work addresses
specific questions related to discovering impacting asteroids
before they hit the Earth.  In particular, we build upon the
work of \citet{Chesley04} and determine the sky-plane distribution
of impacting asteroids before impact and the effectiveness of the
next generation large synoptic sky surveys at identifying
impactors.

The first surveys to target near-Earth objects (NEO) \citep{Helin79},
asteroids and comets with perihelion $<1.3$~AU, provided the first
look at their orbit and size distribution and allowed the first
determination of the impact rate from NEO statistics
\citep{Shoemaker83} rather than crater counting on the Moon.  These
pioneers heightened the awareness of the impact risk and gave rise to
the current generation of CCD-based asteroid and comet surveys such as
Spacewatch \citep{Gehrels86}, LINEAR \citep{Stokes00}, LONEOS
\citep{Koehn99}, NEAT \citep{Pravdo99}, and the current leader in
discovering NEOs, the Catalina Sky Survey (CSS) \citep{Larson98}.
These programs benefitted from the elevated impact risk perception
when in 1998 the U.S. Congress followed the recommendations of
\citet{Morrison92} and mandated that the U.S. National Aeronautics and
Space Administration (NASA) search, find and catalog $\ge 90\%$ of
NEOs with diameters larger than 1~km within 10~years.  That goal will
probably be achieved within the next few years.  The residual impact
risk is mainly due to the remaining undiscovered large asteroids and
comets \citep{Harris07} but \citet{Stokes04} suggest that the search
should be expanded to identify $\ge 90$\% of potentially hazardous
objects (PHO) by 2020.  \footnote{A PHO is an object with absolute
magnitude $H \le 22$ ($\sim 140$~m diameter) on an orbit that comes
within 0.05~AU of the Earth's orbit.}

\citet{Stokes04} showed that the extended goal cannot be achieved in a
reasonable time frame with existing survey technology.  The search
needs to be done from space (rapid completion but at high risk and
high cost) or from new ground-based facilities (slower completion but
lower risk and lower cost).  Their recommendation dovetailed nicely
with the \citet{AA2001} Decadal Report that made a strong case for the
development of a large synoptic survey telescope (LSST) that would provide
the necessary depth and sky coverage to identify the smaller PHOs
while also satisfying the goals of other fields of astronomy.

There are currently a few candidates for a large synoptic survey
telescope.  The most ambitious is an 8.4~m system being designed by
the eponymous LSSTC (the LSST Corporation) that anticipates beginning
survey operations in 2016 in Chile. With a $\sim$9~deg$^2$ field of
view and 15~s exposures, simulations suggest that their system could
identify $\ga$90\% of PHOs in 15 years \citep{Ivezic07}.  A more
modest LSST known as the Panoramic Survey Telescope and Rapid Response
System (Pan-STARRS) will be composed of four 1.8m telescopes (PS4) and
is expected to be located atop Mauna Kea in Hawaii.  A prototype
single telescope for Pan-STARRS known as \psone\ should begin
operations in mid-2009 from Haleakala, Maui.  With a $\sim$7~deg$^2$
field of view, the excellent seeing from the summit of Mauna Kea, and
the use of orthogonal transfer array CCDs \citep{Burke07} for on-chip
image motion compensation, the Pan-STARRS system will be competitive
with and completed earlier than the LSSTC's system.

The next generation survey telescopes have the potential to be
prolific discoverers of PHOs but Earthlings aren't so much concerned
with statistical impact risk calculated from PHO orbital distributions
as they are interested in whether an impact event will occur. The
statistical risk of a house burning down may seem inconsequential
until you consider the actuality of {\it your} house being
incinerated. Similarly, while \citet{Harris08} has calculated that
expected fatalities due to an unanticipated asteroid impact have
dropped from $\sim$1,100/year before the onset of modern NEO surveys
to only $\sim$80/year now, as a species we would like to know whether
one of the fatality inducing impacts will take place {\it this}
century.  Thus, this work concentrates on the detection of objects
that may impact the Earth in the next hundred years.

Following \citet{Chesley04} we concentrate on the subset of PHOs that
are in fact destined for a collision with the Earth.  They showed that
long before impact the impactors' steady state sky-plane distribution
is concentrated on the ecliptic and at small solar elongation.  We
extend their analysis and find that the sky-plane distribution of
impactors has interesting and potentially useful structure in the time
leading to collision.  We also study the capabilities of one
next-generation survey (\psone) at identifying the impactors well
before collision.  In particular, we will answer the following
questions: How different are the orbital characteristics of the
impactor population and current NEO and PHO models?  What is the
survey efficiency for identifying asteroids on a collision course with
the Earth as a function of their diameter?  How much warning time will
be provided before the impact? How accurate is the orbital solution
prior to impact? How does the MOID\footnote{Minimum Orbital
Intersection Distance} evolve in time and is the current definition of
a PHO consistent with flagging dangerous objects? What are the orbital
properties of objects that are not found?  Are there methods to
improve the efficiency of identifying impactors?  Given the
size-frequency distribution of NEOs what is the probability that
\psone\ will actually identify an impactor and what will be its most
probable size?

\section{Synthetic Earth-impacting asteroids}\label{s.Impactors}

Our synthetic impactor population model is described in detail in
\citet{Chesley04} and \citet{Grav09}.  Here we provide a brief summary
of the technique.  

We created $\sim$130,000 impactors based on the NEO population
developed by \citet{Bottke00} and \citet{Bottke02} hereafter referred
to as the Bottke NEO Model.  The model incorporates objects from both
asteroidal and cometary source regions but has at least two problems
that affect its utility for creating an impactor population: 1) it
assumes that the orbit distribution of NEOs is independent of their
diameter and 2) it provides the $(a,e,i,H)$ (semi-major axis,
eccentricity, inclination, and absolute magnitude) distribution for
NEOs on a coarse grid that is not suited to the narrower range of
orbital elements of the impacting asteroids. However, there are few
options to use as starting points for developing an impactor
population and we will compare our impactor population's orbit
distribution to the known small impactor population to understand the
limitations of our technique.

To generate the impactors NEOs were randomly selected from the Bottke
NEO model and assigned random longitudes of ascending node and
arguments of perihelion. Orbits with a MOID small enough to permit an
impact were saved as {\em potential impactors} and then filtered
according to their likelihood of impact to obtain the final set of
impactors.  The likelihood is the fraction of time that an object
spends in close proximity to the Earth's orbit. \ie orbits with a
small velocity relative to the Earth tend to have shorter impact
intervals and higher intrinsic impact probabilities. Higher
likelihoods received higher weighting in the selection.  If an orbit
was chosen as an impactor then a year of impact was randomly selected
between 2010 and 2110 --- the date of collision is already randomly
fixed by the longitude of the node at impact. To this point, the
process assumed a two-body asteroid orbit with no planetary
perturbations. The final step was to ensure an impact under the
influence of all the perturbations in a complete solar system
dynamical model. This was done by differentially adjusting the
two-body argument of perihelion ($\omega$) and orbital anomaly to
reach a randomly selected target plane coordinate on the figure of the
Earth. The final result is an osculating element set that leads to an
Earth impact when propagated with the full dynamical model.  The full
set of impactors generate about three impacts per day uniformly
distributed over the globe with an average separation of about 70
km.\footnote{\citet{Gallant06} use a superior (but much more time
consuming) technique to generate an even more unbiased impactor
population from the Bottke NEO model and confirm that the latitude and
longitude distribution of impact locations is flat to within a few
percent when averaged over all impactors and times of year.}

This technique preferentially selects objects on Earth-like orbits out
of the Bottke NEO model but \citet{Brasser08} show that objects do not
remain long in these types of orbits.  This is not a problem except in
the sense addressed above --- that the Bottke NEO model is provided on
a relatively coarse grid --- because the NEO model already accounts
for NEO `residence times' on all types of NEO orbits.  However, since
we assume a flat distribution of NEO orbit elements within the
$(a,e,i)$ bin corresponding to Earth-like orbits it is likely that we
generate fractionally more of the extremely Earth-like orbits than
exist in reality.

\begin{figure}[htp]
\centering
\includegraphics[width=0.8\textwidth]{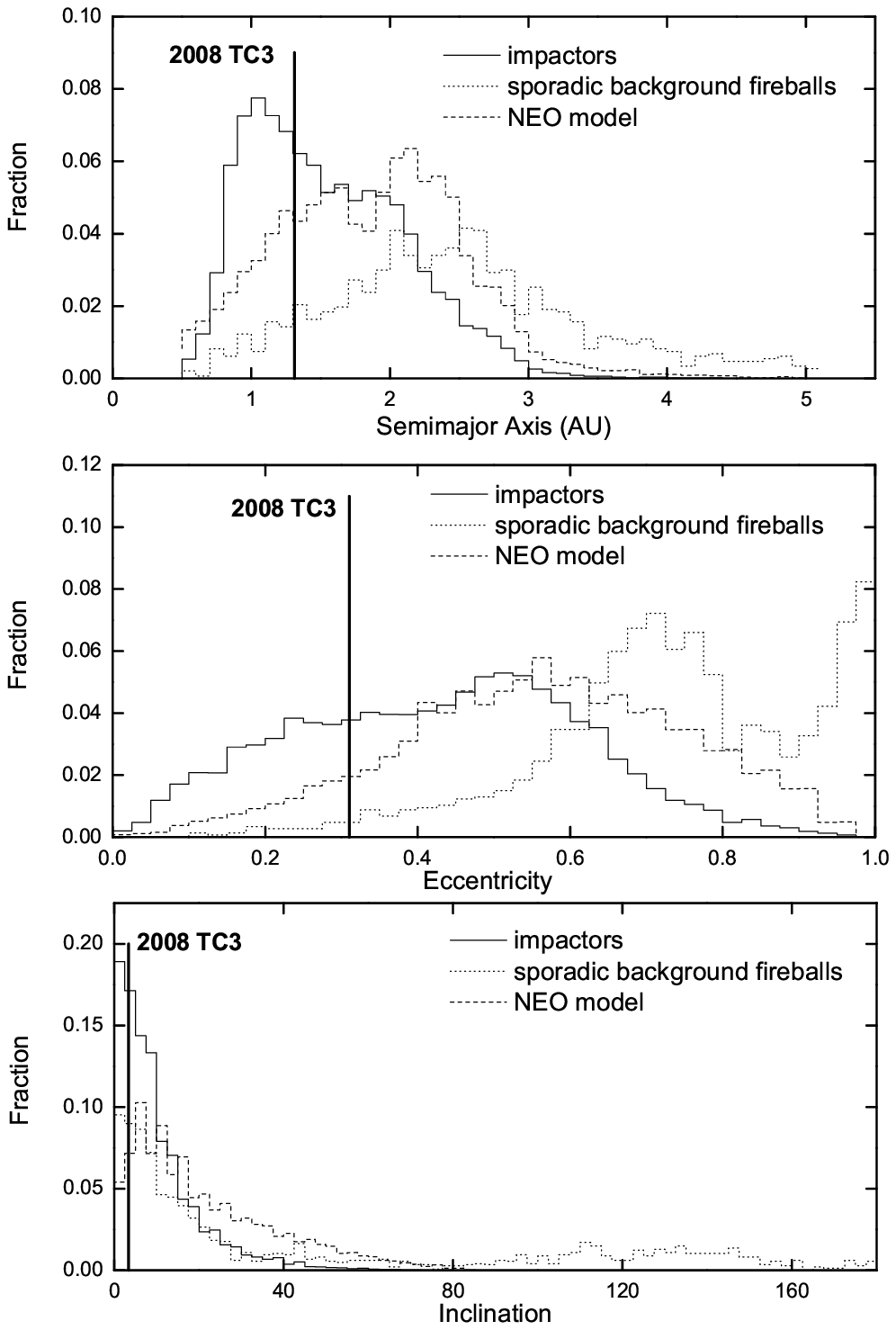}

\caption{Semi-major axis ({\it{top})}, eccentricity ({\it{middle}})
and inclination ({\it{bottom}}) distribution for ({\it solid}) our
synthetic impactor population, ({\it dotted}) sporadic background
fireballs and ({\it dashed}) the Bottke NEO model.  The
semi-major axis distribution is not shown beyond 5.1~AU but these data
are included in the eccentricity and inclination distributions. The
corresponding orbital elements for 2008~TC3 are indicated by the solid
vertical line in each figure.}

\label{fig.ImpactorBolideComparison}
\end{figure}

As shown by \cite{Chesley04} and in
Fig. \ref{fig.ImpactorBolideComparison} there are important
differences between the impactor population and the NEOs.  The
impactors have orbits with lower semi-major axis, inclination and
eccentricity.  This has the effect of decreasing the Earth encounter
and impact velocity ($v_\infty$ and $v_{imp}$ respectively) for the
impactors relative to the NEOs.  The decreased impact velocity has
implications for modelling the impact crater size-frequency
distribution on the Earth and Moon since lower impact energies per
unit mass require larger impactors to create equivalent-size craters.

As mentioned above, the Bottke NEO model assumed that the orbit
distribution of the NEOs is independent of size.  While this
assumption is probably fine for large objects it must break down at
smaller sizes due to the effect of non-gravitational forces such as
the Yarkovsky effect \citep[\eg][]{Bottke98,Obrien05}.  Where the
transition occurs is not clear but
Fig. \ref{fig.ImpactorBolideComparison} compares the $(a,e,i)$
distribution of our impactor population to sporadic fireballs from
the IAU Meteor Database of photographic orbits \citep{Lindblad03}. We
removed fireballs due to 17 major meteor showers\footnote{From the IAU
Meteor Database: Quadrantids, Lyrids, Pi Puppids, Eta Aquarids,
Arietids, Daytime Zeta Perseids, June Bootids, Southern Delta
Aquarids, Perseids, Draconids, Orionids, Southern Taurids, Northern
Taurids, Leonids, Puppid/Velids, Geminids, Ursids.}
\citep{Jenniskens06} by requiring that the meteors not be associated
with the parent meteor body using the dimensionless orbit similarity
$D$-criterion of \citet{Valsecchi99}\footnote{We obtain essentially
identical results using other $D$-criterion formulations
\citep{Southworth63,Drummond81,Galligan01} with corresponding but
different upper limits on the $D$-criterion value.}:
\begin{equation}\label{eq.Dcriterion}
D^{2} = [U_{2} -U_{1}]^{2} + w_{1}[\cos \theta_{2} - \cos \theta_{1}]^{2} + 
\Delta \xi^{2},
\end{equation}
where
\begin{eqnarray}
\nonumber \cos \theta &=& \frac{1-U^{2}-1/a}{2U},\\
\nonumber \Delta \xi^{2} &=& \min \bigl[ \; w_{2} \Delta \phi^{2}_{A}
  + w_{3} \Delta \lambda^{2}_{\oplus A}, \; w_{2} \Delta \phi^{2}_{B} + w_{3} \Delta \lambda^{2}_{\oplus B}\; \bigr],\\
\nonumber \phi_{A} &=& 2 \sin \Bigl(\frac{\phi_{2} - \phi_{1}}{2}\Bigr),\\
\nonumber \phi_{B} &=& 2 \sin \Bigl(\frac{\pi + \phi_{2} - \phi_{1}}{2}\Bigr),\\
\nonumber \lambda_{\oplus A} &=& 2 \sin \Bigl(\frac{\lambda_{\oplus 2} - \lambda_{\oplus 1}}{2}\Bigr),\\
\nonumber \lambda_{\oplus B} &=&  2 \sin \Bigl(\frac{\pi + \lambda_{\oplus 2} - \lambda_{\oplus 1}}{2}\Bigr).
\end{eqnarray}
The 1 and 2 subscripts refer to the two bodies whose orbits are being
compared, U is the unperturbed geocentric speed just prior to impact,
($\theta$, $\phi$) define the direction of the radiant in a frame
moving with the Earth about the Sun, and $\lambda_\oplus$ is the ecliptic
longitude of the Earth at the time of meteoroid impact. The weighting
factors $w_{i}$ were set to 1 as per \citet{Valsecchi99}. All objects
with $D$-criterion relative to a parent body of $\leq 0.2$ were
discarded (577 meteors) leaving 2002 sporadic background fireballs.

The impactor and NEO orbit distributions have already been discussed
briefly above and in detail by \citet{Chesley04}.  The differences
between the impactor population and the fireballs are perhaps more
interesting where it is important to keep in mind that the comparison
in Fig.  \ref{fig.ImpactorBolideComparison} is between the {\it bias
corrected} impactor population and the {\it observed} bolide
population. We believe that the apparent difference between the
impactor and bolide distributions is a consequence of the uncorrected
observational selection effects in the bolide data along with
modifications in the distributions due to the Yarkovsky effect
\citep[\eg][]{Farinella98}.  The bolide detection technique has a
strong bias towards high kinetic energy events that preferentially
detects objects on cometary-like orbits \citep{Ceplecha98}.  Indeed,
if we consider `comets' to be objects from our sporadic bolide data
with $a>4$~AU or $e>0.9$ or $i>90\deg$ then $\sim$13.4\% of the
objects are of `cometary' origin.  This value is about twice the
cometary fraction of $6\pm4$\% suggested by the Bottke NEO model and
consistent with the expected `comet' enhancement in the bolide data.
On the other hand, it is only half the 25\% cometary contribution
suggested by \citet{Stuart04}.

No meteor or fireball had ever been detected before entering the
Earth's atmosphere before we generated our synthetic population of
Earth-impacting asteroids\footnote{There were suspected radar
detections of exoatmospheric meteoroids in the late 1970's
\citep{Kessler80}}. Then, on 7 October 2008, asteroid 2008~TC$_3$ was
discovered by CSS using their 1.5~m telescope.  Rapid followup by
other observatories made it almost immediately clear that the
few-meter-diameter ($H=30.7$) asteroid would enter the Earth's
atmosphere within a day and explode over northern Sudan. The
substantial and largely self-organized follow-up effort resulted in
789 astrometric observations from amateur and professional
observatories worldwide being submitted to the Minor Planet
Center. Due to the parallax induced by the distance between the
observatories and the proximity of the asteroid, the observations
allowed an accurate pre-impact orbit to be computed despite the short
observational timespan (Table~\ref{tab.2008TC3Orbit}). The pre-impact
orbit is very consistent with our predicted impactor orbit
distribution as shown in Fig.~\ref{fig.ImpactorBolideComparison}. Note
that, in particular for the eccentricity, the pre-impact orbit matches
the expected impactor population better than the debiased NEO
population or the observed bolide population.

\begin{table}[h!]
  \def~{\hphantom{0}}
  \centering
  \begin{tabular}{lrrrrrr}
    \multicolumn{7}{c}{Orbital elements and their uncertainties for 2008~TC$_3$} \\
    \hline
    & $a$ [AU]  &  $e$ & $i$ [$\deg$] & $\Omega$ [$\deg$] & $\omega$ [$\deg$] & $M_0$ [$\deg$] \\
    \hline
    Elements       & 1.2712175 & 0.2856863 & 2.331633 & 194.1308964 & 233.954719 & 328.58963 \\
    1-$\sigma$ unc & 0.0000031 & 0.0000023 & 0.000017 &   0.0000011 &   0.000040 &   0.00015 \\
    \hline
  \end{tabular}

\caption{Keplerian elements and their 1-$\sigma$ uncertainties for
2008 TC$_3$ for the epoch 2008 Oct 6.11535 TT, about one day before
atmospheric entry on 2008 Oct 7.1146 UTC. The orbital solution made
use of 574 astrometric observations (42 of which were discarded as
outliers) and assumed an uncorrelated astrometric uncertainty of 0.5
arcseconds for every measurement. The orbital solution was obtained
using the OpenOrb software \citep{Granvik08}.}

\label{tab.2008TC3Orbit}
\end{table}

\subsection{Time evolution of the impactor population's MOIDs}

\begin{figure}[htp]
\centering
\includegraphics[width=0.7\textwidth]{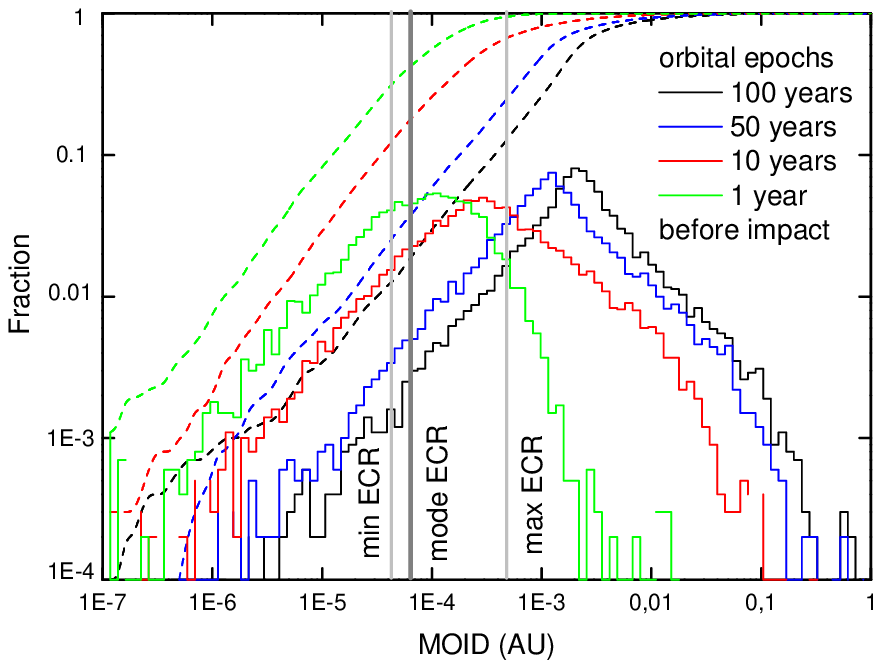}
\includegraphics[width=0.7\textwidth]{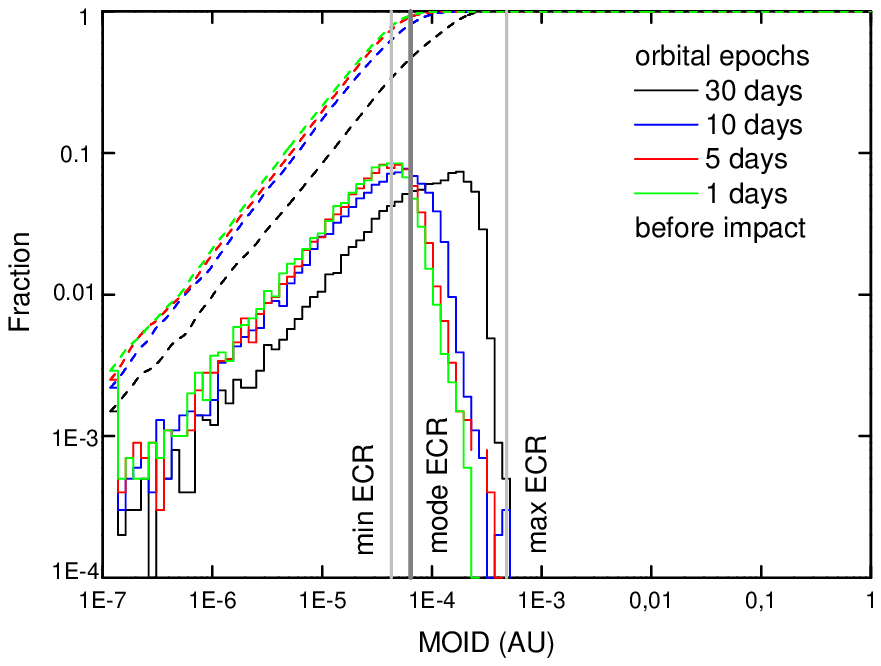}

\caption{Fractional differential ({\it solid}) and cumulative ({\it
dashed}) distribution of MOID for synthetic impactors at different
times before impact.  The Earth capture radius (ECR) is calculated for
the highest (max ECR), most probable (mode ECR) and lowest (min ECR)
impactor encounter velocity.  The bin size is $\log(MOID/AU)=0.07$ in
both figures.}

\label{fig.MOID}
\end{figure}

\begin{figure}[htp]
\centering
\includegraphics[height=120mm, width=150mm]{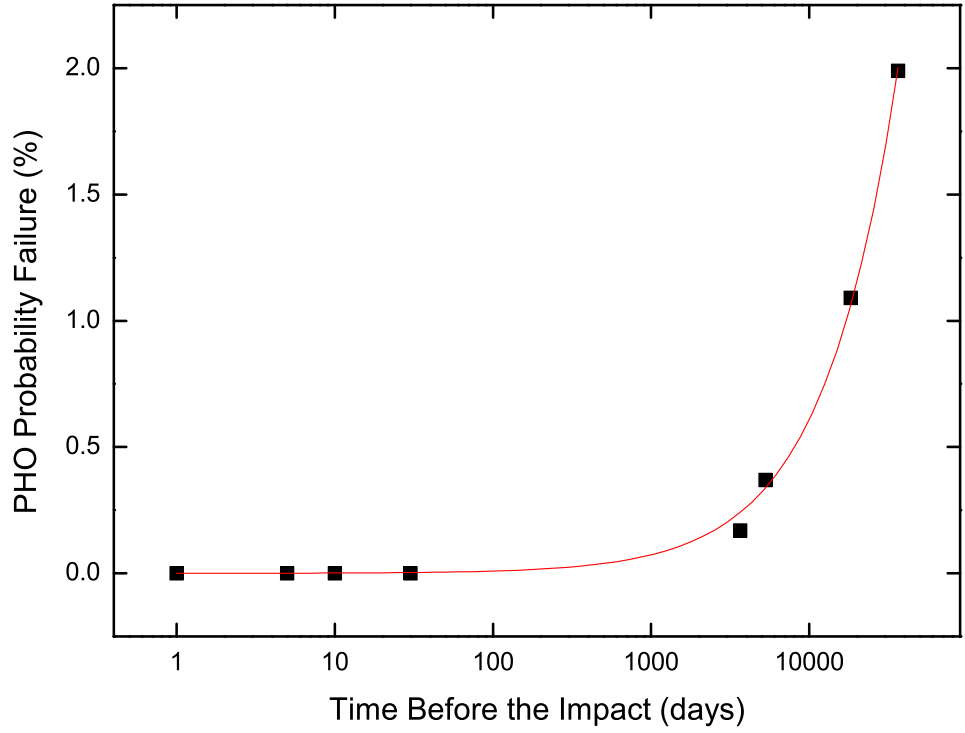}

\caption{The probability that an Earth-impacting asteroid will not be
identified as a PHO (MOID$>0.05 AU$) as a function of time before
impact.  The superimposed line is a fit ($R^2=0.99$) to an empirically
defined function: $p = 1.3 \times 10^{-4} + t^{0.92}$.}

\label{fig.PHOprobability.vs.time}
\end{figure}

Recall that a PHO is defined as an object with $H \le 22$ and a
MOID~$<0.05$~AU with respect to the Earth's orbit.  We investigated
the evolution of the MOID for our impactor population as a function of
time before impact as shown in Fig.~\ref{fig.MOID}.  All the
impactors's osculating orbits were integrated using the JPL's N-body
integrator incorporating the effects of the Sun, eight planets, and
the dwarf planets Pluto, Ceres, Vesta and Pallas.  MOIDs were then
calculated at different times before impact using the synthetic
objects's integrated osculating orbits at the time of interest, not the
orbit derived from synthetic observations of the object.  We see that
as the time of impact approaches the MOID decreases such that one
month before impact essentially all objects have a MOID less than the
Earth's capture radius ($b$):
\begin{equation}\label{eq.earthCaptureRadius}
{b = {R\cdot\sqrt{{1+\frac{v_e^2}{v_\infty^2}}}}},
\end{equation}
where $v_{e}$ represents the escape velocity from the surface of the
Earth.  Fig. \ref{fig.PHOprobability.vs.time} shows that $\sim$99\% of
all impactors are identified as PHOs 45 years in advance of impact.
Even 100 years before impact $\sim$98\% of the objects have a
MOID$<$0.05~AU though the fraction of non-PHO impactors is increasing
rapidly as the time before impact increases.

There are a few objects with MOID$>$0.05~AU that would not be
identified as impactors or PHOs using a single MOID determination from
a derived osculating orbit at the time of discovery.  These objects
suffer from a close approach to Jupiter that converts them from a
harmless object into an Earth-impactor.  Modern impact monitoring
sites such as JPL \citep{Milani05} and NEODyS \citep{Chesley99}
integrate the orbits of all objects to identify these unusual but
dangerous cases.

\begin{figure}[htp]
\centering
\includegraphics[height=80mm, width=100mm]{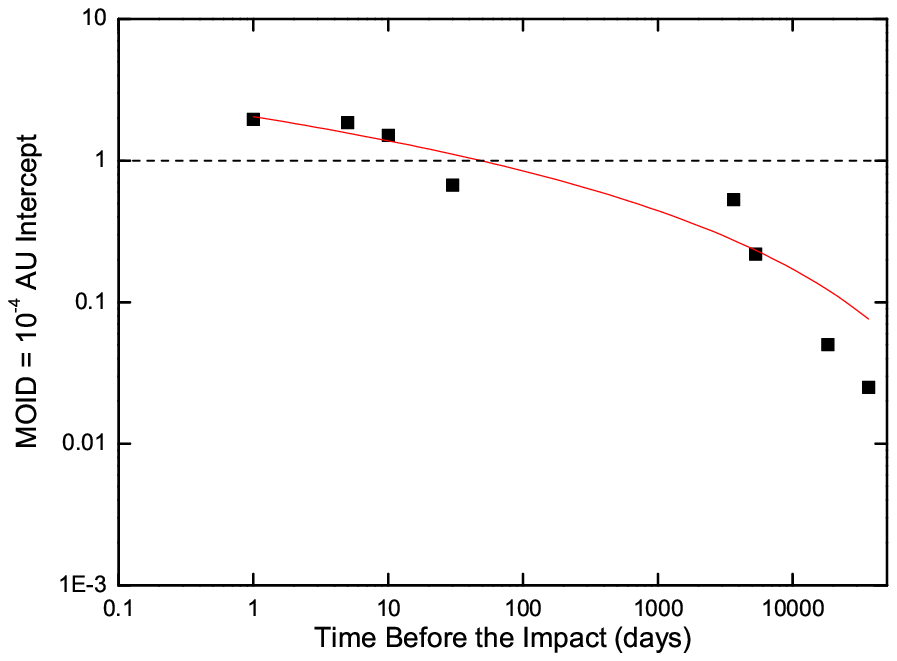}
\includegraphics[height=80mm, width=100mm]{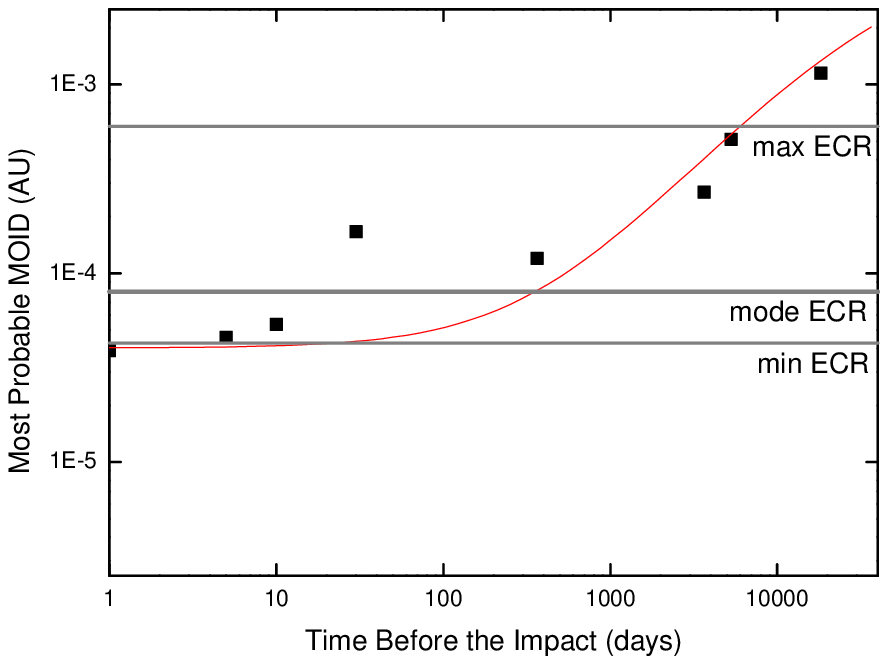}

\caption{ ({{\it top}}) The constant ($f^\prime$) in a fit ($\log f =
\log f^\prime + m \log[MOID/(10^{-4} AU)]$ with $f<0.9$) to the
cumulative fraction ($f$) of impactors with MOID as a function of time
before impact from Figure \ref{fig.MOID}.  \ie $f^\prime$ is the
cumulative fraction of objects with MOID$\le 10^{-4}$~AU at each of
the specified time steps.  The superimposed solid line is a fit
($R^2=0.954$) to an empirically defined function: $\log f^\prime = a
[\log^2 (t/days) - \log^2 (t_0/days)]$ with $t$ being the time before
impact, $a=-0.079$ and $t_0=72.8$~days.  ({\it bottom}) The most
probable MOID (mode) for the impactors as a function of time before
impact.  The superimposed solid line is a fit ($R^2=0.90$) to an
empirically defined function: MOID (AU)$ = -0.014 - 1.5 \times 10^{-3}
\ln{[((t/days)+10^{5})]}$. The Earth capture radius is shown for the
highest (max ECR), most probable (mode ECR) and lowest (min ECR)
encounter velocity for the impactors.  }

\label{fig.MOIDSlopeMode.vs.time}
\end{figure}

At all eight times before impact the cumulative fractional distribution
of MOIDs ($f_C$) in Figs.~\ref{fig.MOID} exhibit a nearly constant
slope for $f_C \la 0.9$.  Under the simple assumption that the
impactors are randomly distributed on the impact plane at the Earth we
would expect $f_C \propto MOID^2$.  Thus, we fit the distribution to
$\log f_C = \log f_C^\prime + m \log[MOID/(10^{-4} AU)]$ where
$f_C^\prime$ is (roughly) the cumulative fraction of objects with
$MOID<10^{-4}$~AU.  We find an average slope over all impact times of
$m = 0.88 \pm 0.01$, much less than 2 and quite different from the
result of \citet{Tancredi98} who found a slope of $\sim$1.23 for
clones of the extremely Earth orbit-like 1991 VG.  We believe that the
difference between the expected quadratic and our measured unit slope
is due to the Earth's gravitational focussing.  The cumulative
fraction of impactors with MOID$<10^{-4}$~AU as a function of time is
shown in Fig~\ref{fig.MOIDSlopeMode.vs.time}a.  Clearly, the
interpration of $f_C$ as a cumulative fraction breaks down for $f_C
\ga 1$ but we see that essentially all the impactors have
MOID$<10^{-4}$~AU about 73 days before
impact. Fig~\ref{fig.MOIDSlopeMode.vs.time}b provides a quantitative
assessment of the time evolution of the mode of the MOID distribution
from Fig.~\ref{fig.MOID}.  The combination of the fits to the data in
Figures~\ref{fig.MOIDSlopeMode.vs.time} allow a rough determination of
the time evolution of the fraction of impactors with a given MOID.

\subsection{Sky-plane distribution of Earth-impacting asteroids}
\label{s.SkyPlaneDistribution}

In this section we analyze the sky-plane distribution of
Earth-impacting asteroids in the years, months, and days leading up to
impact.  The topocentric location and brightness of the impactors (as
seen from Mauna Kea in all cases) was calculated for a $\sim$10K
subset of impactors for every day in the 100 years leading to impact
using the OpenOrb software package \citep{Granvik08}.

\begin{figure}[htp]
\centering
\includegraphics[width=0.9\textwidth]{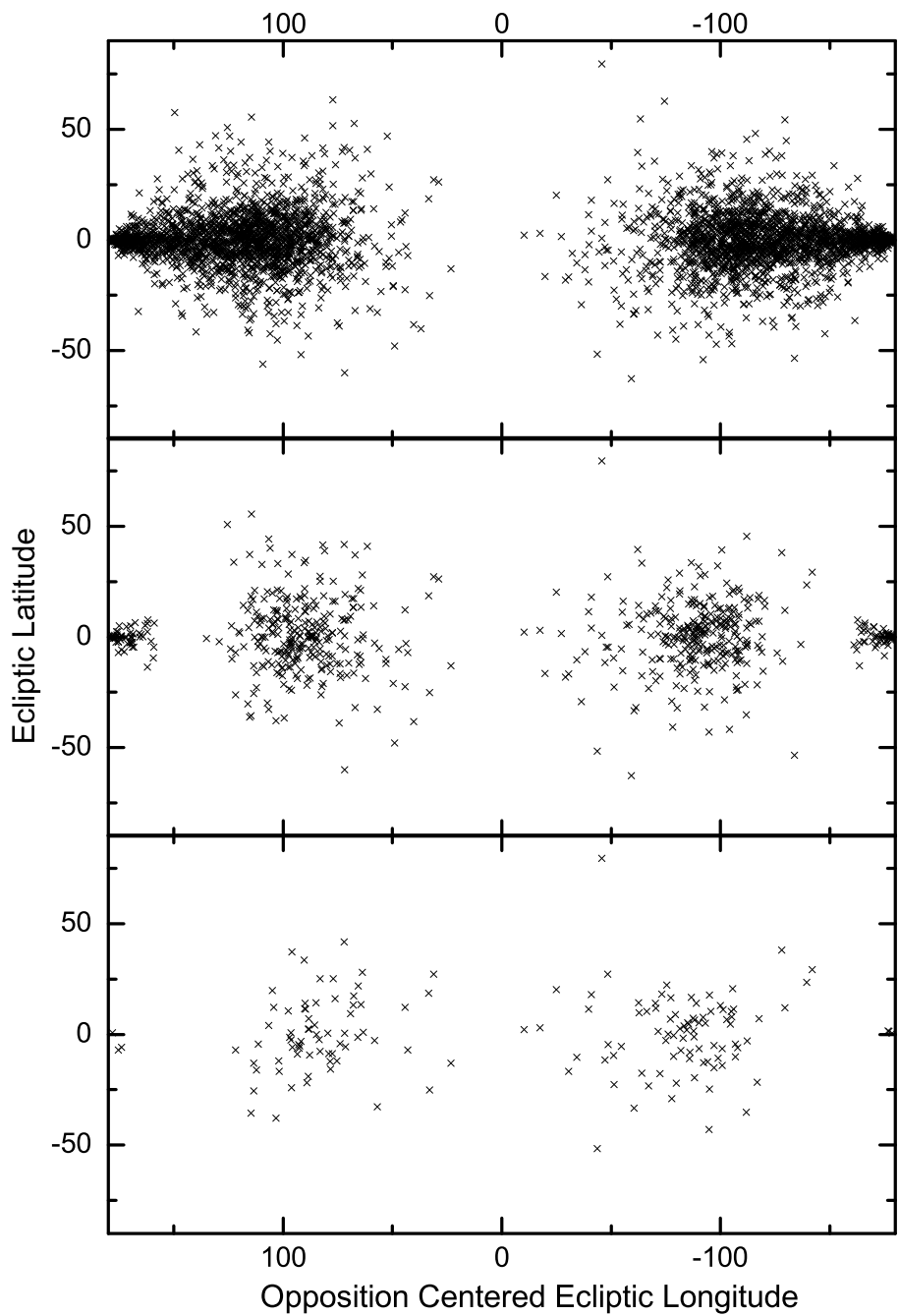}

\caption{Topocentric opposition centered ecliptic sky plane
distribution of $H=20$ ($\sim$350~m diameter) Earth-impacting
asteroids 20 years before impact for a survey with a limiting
magnitude of ({\it top}) $V=22.7$ ({\it middle}) $V=20.7$ and ({\it
bottom}) $V=18.7$.  This is sufficiently in advance of impact to
represent the impactors's steady-state sky plane distribution.}

\label{fig.20years}
\end{figure}

The steady-state\footnote{Although Figure~\ref{fig.20years}a shows the
sky-plane distribution of Earth-impacting asteroids 20 years before
impact we have verified that the distribution is the same for earlier
pre-impact times.} sky plane distribution of Earth-impactors in
Figure~\ref{fig.20years}a shows all objects that would be visible to
PS1 with $V<22.7$ assuming that $H=20$ (diameter $\sim$350~m).  The
figure clearly shows the high sky-plane density `sweet spots' at small
solar elongations ($\pm \sim 90\deg$) identified by \citet{Chesley04}.
The objects in the sweet spots are close to the Earth and therefore
relatively bright despite their large phase angles, but are too close
to the Sun to be easily observed.

\begin{figure}[htp]
\centering
\includegraphics[width=1.0\textwidth]{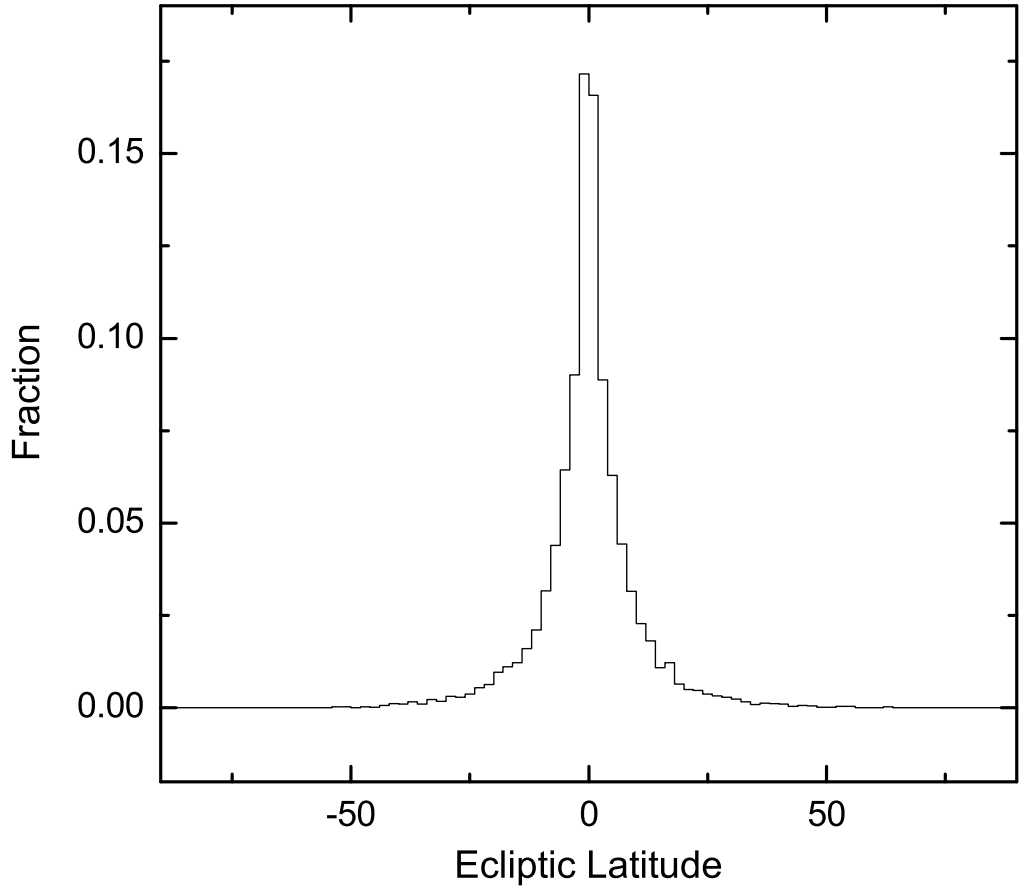}

\caption{Ecliptic latitude distribution of Earth-impacting asteroids
 with $V<22.7$ twenty years before impact.}

\label{fig.20years-latitude}
\end{figure}

Twenty years before impact 962 objects ($\sim$9.6\%) lie within the
sweet spot region that encompasses topocentric solar elongations from
$60\deg$ to $90\deg$.  Of these, 552 ($\sim$5.5\%) have an ecliptic
latitude ($\beta$) in the range $-10\deg \le \beta \le +10\deg$ and
826 ($\sim$8.3\%) have $|\beta|<20\deg$. The distribution in ecliptic
latitude (Fig.~\ref{fig.20years-latitude}) is distinctly broader than
provided by \citet{Chesley04} who showed cumulative detections made by
a simulated version of the LINEAR survey \citep{Stokes00}.  The broad
distribution in ecliptic latitude suggests that there is merit in
extending searches for PHOs in the sweet spots to ecliptic latitudes
of $\pm 20^\circ$.  Since the southern parts of these extended sweet
spots are close to the horizon from the northern hemisphere, and the
northern parts of the extended sweet spots are close to the horizon
from the southern hemisphere, this in turn suggests that there is
merit in conducting searches for PHOs from both hemispheres or from
space \citep[\eg][]{Hildebrand07}.

The middle and bottom distributions in Fig.~\ref{fig.20years} show the
location of the same sample of objects 20 years before impact but with
brighter magnitude limits of $V<20.7$ and $V<18.7$.  Another way to
interpret these figures is that they show the detectability (for the
same limiting magnitude of $V<22.7$) of $H=22$ (diameter 140~m) and
$H=24$ (55~m) impactors at this particular time.  The smaller objects
must be closer to the Earth in order to be above the system's limiting
magnitude but this also has the effect of increasing the phase angle
which further decreases the apparent magnitude.  Thus, smaller objects
must be closer {\it and} have smaller phase angle to be detected.
While searches for larger objects are efficient in the sweet spots
they should be supplemented by opposition surveys to find smaller
objects.  \eg 20 years before impact 2008~TC$_3$ had an apparent
magnitude of $V \approx 31$ with an ecliptic opposition-centered
longitude of $\lambda_{opp} \approx -92^\circ$ and latitude of $\beta
\approx 2^\circ$ --- not detectable but in the sweet spot region.

\begin{figure}[htp]
\centering
\includegraphics[width=0.9\textwidth]{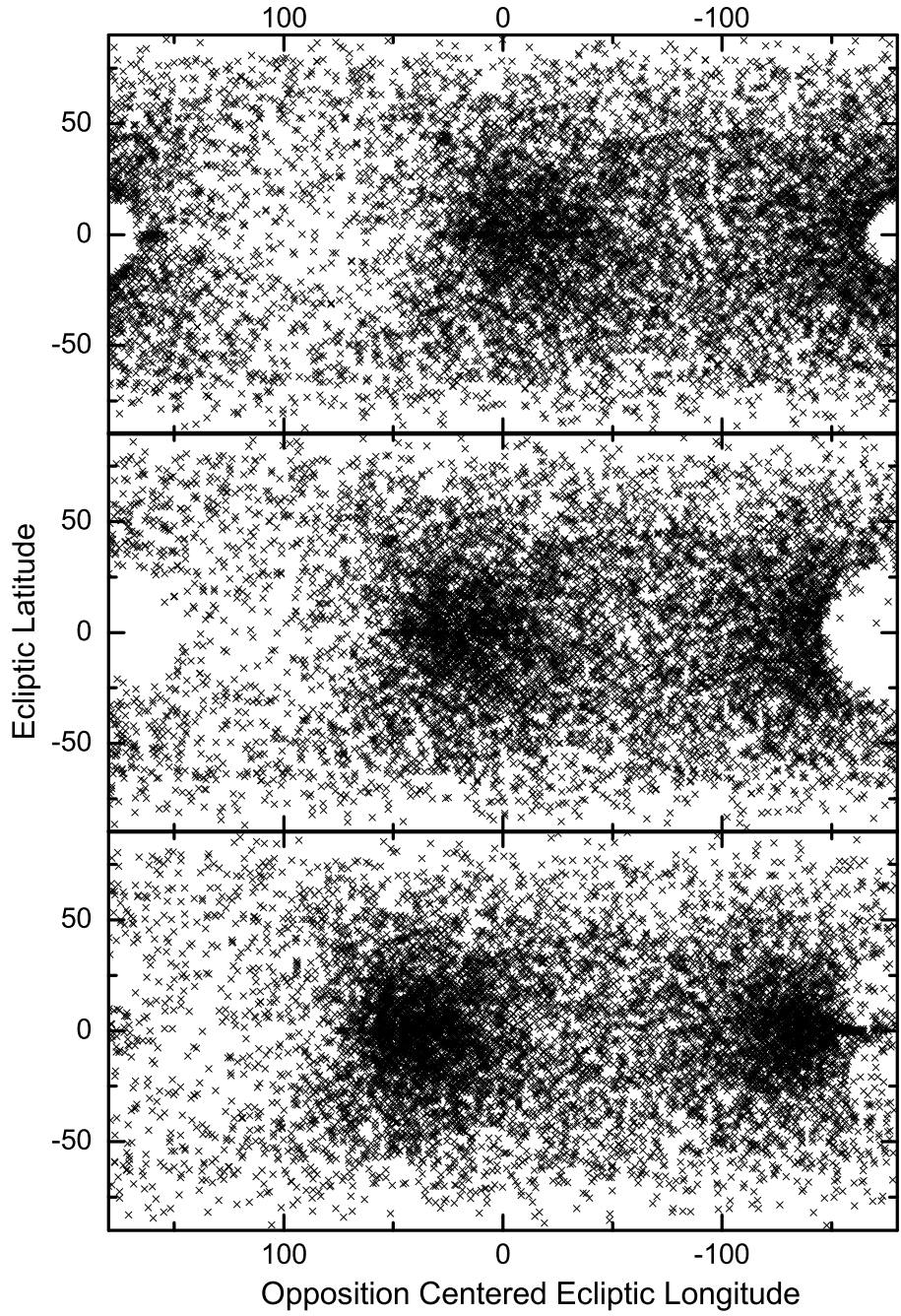}

\caption{a) Topocentric opposition centered ecliptic sky plane
distribution of Earth-impacting asteroids with $V \le 22.7$ ({\it top})
1 day, ({\it middle}) 30 days, and ({\it bottom}) 60 days before
impact.}

\label{fig.skyplaneTimeEvolution-a}
\end{figure}

Figures~\ref{fig.skyplaneTimeEvolution-a}a-d show the development of
the sky-plane distribution of impactors as a function of time before
impact.
\begin{itemize}
\item 1 day before impact the impactors are located in two main
  concentrations; one centered on the opposition direction and the
  other centered on the Sun.  The impactors are widely spread in
  ecliptic latitude because they are close to the Earth.  About half
  as many impactors approach from the $\lambda_{opp} > 0$
  (a.m./morning) side compared to the $\lambda_{opp} < 0$
  (p.m./evening) side.  Objects that approach from the morning side
  are moving more slowly around the Sun than the Earth, meaning that
  the Earth runs into them.  They have perihelia that are well inside
  the Earth's orbit, or have higher inclinations.  Objects that impact
  on the evening side catch up with the Earth in its orbit around the
  Sun.  They have perihelia that are closer to the Earth's orbit, or
  have higher eccentricities.
\item 30 days before impact, the concentration that will impact from
  the opposition direction has moved east and is centered close to
  $\lambda_{opp}=20^\circ$.  The impactors that will approach from close to
  the Sun direction are also further east with some of them becoming
  observable in the evening sweet spot.
\item 60 days before impact many of the impactors that will approach
  the Earth from outside the Earth's orbit are scattered around
  $\lambda_{opp}=40^\circ$ and more of the impactors that will approach
  from inside the Earth's orbit are becoming visible near the evening
  sweet spot.  This trend continues 90 days and 120 days before
  impact with the outside impactors moving further east and more of
  the inside impactors becoming more visible in the evening sweet
  spot.
\item More than 180 days before impact the sky plane distribution of
  the impactors is more complex.  The observability of impactors
  decreases significantly because of their small solar elongations.
  180 days, 1 year and 1.5 years before impact many impactors are
  close to the direction of the Sun or are far from the Earth.
  Observability improves from 2 years onwards as the distribution more
  closely mimics the steady state distribution of impactors in the
  sky.
\end{itemize}

The difficulty in observing many of the impactors in the period
between 0.5 and 1.5 years before collision with the Earth is
important.  In the event that an impact is predicted, precision
astrometry will be needed during this period to determine where and
when the impact will occur so that life-saving actions can be taken if
necessary.

\section{Next generation sky survey impactor detection performance}
\label{s.SurveySimulation}

Our goals in this section are to determine
\begin{itemize}
\item the efficiency of a next-generation sky survey at identifying
  Earth-impacting asteroids
\item the likely warning time for impending impacts
\item the reason(s) why some impactors remain undetected
\item methods for discovering the impactors that are most difficult to
  detect
\item the probability that a next-generation survey will
  identify an impacting object as a function of its diameter
\end{itemize}

There have been numerous efforts in the recent past to simulate the
performance of individual ground and/or space-based sky surveys at
detecting NEOs
\citep[\eg][]{Bowell94,Jedicke03,Harris04,Stokes04,Ivezic07,Moon08}
but only \citet{Chesley04} concentrated specifically on identifying
impactors and their work also simulated the performance of the last
generation of asteroid surveys.  As discussed above, the impactors
have a different orbit distribution from the NEO or even PHO
population and pose different challenges to a survey and its moving
object detection system.  These include some pathological cases in the
distribution of synodic periods for the objects, the fact that they
are infrequently close enough and above the limiting magnitude of the
detection system, and that at small topocentric distance their
apparent rate of motion on the sky can be very small (mimicking more
distant objects) and rapidly varying due to the topocentric motion of
the observer. For all these reasons, when simulating the detection of
impactors it is probably not sufficient to assume that all the objects
will be detected and identified as imminently hazardous.  A
high-fidelity simulation is required that models the entire detection
process from imaging through orbit determination and hazard
assessment.

The Pan-STARRS surveying mode has not yet been decided upon and,
realistically, will asymptotically approach its final configuration
during the first year of operation.  On average, each month \psone\
will survey $\pi$/2 steradians or $\sim$5,000 square degrees in an
`opposition' survey with an overlap between months of about $\pi$/4
steradians. The opposition fields are imaged in g, r, and i each
lunation near new moon. The redder (z, y) filter observations are
obtained near quadrature closer to full moon to take advantage of
their reduced sensitivity to scattered moonlight.  Each field will be
imaged twice (15-30 minutes apart) in each filter in each month and
imaged in two lunations/year due to the overlapping survey area from
month to month.

We employed the Pan-STARRS Moving Object Processing System (MOPS) to
process synthetic detections generated in a pseudo-realistic
Pan-STARRS survey.  The simulated survey covers a large region
(3600~deg$^2$) of the sky around opposition (within roughly
$\pm30\deg$ from opposition in longitude and $\pm30\deg$ from the
ecliptic) and two $\sim 600$~deg$^2$ regions in the sweet spots
defined by \citet{Chesley04} within $\pm 10\deg$ of the ecliptic and
from about $60\deg$ to $90\deg$ solar elongation.  It is essentially a
solar system specific sub-set of the sky that the Pan-STARRS system is
expected to cover when it becomes operational.

The survey simulator is described in detail in \citet{Jedicke05} and
\citet{Denneau07}.  It incorporates a crude weather simulation and a
realistic survey pattern that attempts to observe fields at high
altitude and on the meridian. We use a full N-body ephemeris
determination to calculate the exact (RA,Dec) of each impactor in each
synthetic field and then add noise to the astrometric position
according to the expected \psone\ S/N-dependent astrometric error
model. The photometry for each object is similarly degraded and then
we make a cut at S/N=5 in order to simulate the statistical loss of
detections near the system's limiting magnitude ($V=22.7$). Each field
is observed twice each night within $\sim$15~minutes to allow the
formation of `tracklets' --- pairs of detections at nearly the same
spatial location that might represent the same solar system
object. Fields are re-observed 3 times per lunation (weather
permitting) and tracklets are linked across nights to form `tracks'
that are then tested for consistency using an initial orbit
determination (IOD).  Detections in tracks with small astrometric
residuals in the IOD are subsequently differentially corrected to
obtain a final orbit.  The major item that is not simulated is the
effect of the camera fill-factor and this could have a major impact on
the survey's efficiency.  We will estimate this effect below but the
simulation is otherwise one of the highest fidelity moving object
simulations ever attempted.

We anticipate that there may be concerns regarding simulating the
performance of detector systems especially in the sense that the
simulations tend to be optimistic compared to the actual detector.  In
particular, in this case we have
\begin{enumerate}
\item used a realistic survey scenario but it is not the survey that
      will eventually be implemented by \psone,
\item used a simplistic weather model,
\item used a S/N-dependent astrometric error that is more appropriate
      to future surveys with better catalogs (at the onset of
      operations \psone\ will probably use the USNO-B catalog
      \citep{Monet2003} which will limit absolute astrometry to worse
      than 0.1\arcsec),
\item not incorporated false detections,
\item used an early version of the MOPS with reduced efficiency
      compared to the most recent version ($\sim$80\%
      vs. $\sim$100\%),
\item not accounted for the camera fill-factor (the fraction of `live'
      pixels on the detector compared to the footprint of the focal
      plane on the sky).
\end{enumerate}

We think that the first three factors are relatively unimportant.  The
implemented survey scenario is quite good and surveys most of the sky
in which solar system objects will appear.  The weather model is
simple but has the desired effect of disrupting the cadence of
observations and sometimes eliminating lunations from consideration
due to there being too many bad nights.  We have also tested MOPS
performance under conditions of larger astrometric uncertainty and
find that it still performs well.

The fact that no false detections were used in the simulation is also
unimportant.  In targetted smaller simulations \citep[\eg][]{Milani08}
we tested the MOPS using a full density complement of false and
synthetic asteroid detections and found no degradation in performance
at the 5-$\sigma$ level.  In those tests the minimum ratio of
false:synthetic detections was 1:1 on the ecliptic where the density
of synthetic asteroid detections is highest.  Since the density of
asteroids decreases rapidly with latitude the false:synthetic ratio
increases dramatically toward the ecliptic poles.  Still, few
tracklets incorporate false detections because those detections are
spatially uncorrelated (in the simulation).  It is likely that real
survey systems will produce correlated false detections near chip
gaps, due to CCD defects, diffraction spikes, etc., and this will
produce more false tracklets than our simulation.  However, the false
tracklets will not link to other tracklets on other days and, if they
do, will not pass quality control checks on the derived orbit.
Furthermore, MOPS makes use of the detection's morphology when
combining detections into tracklets - the distance between the
detections must be consistent with trailing observed in the detections
themselves.  Indeed, we have employed MOPS to identify asteroids in
real data from the CFHT telescope \citep{Masiero09} and from
Spacewatch \citep[\eg][]{Larsen2007} with excellent performance.

The last two factors are most important.  The reduced efficiency of
the MOPS software used in this simulation will have a proportional
effect on the number of simulated impactor discoveries - about a 20\%
reduction.  The actual system performance will therefore be about 25\%
{\it better} than reported here.  

On the other hand, the largest negative impact on the simulation will
be the \psone\ camera fill-factor. The effective camera fill-factor
($f$) is reduced by the metal lines between individual OTA cells, the
gaps between the CCDs, the use of some cells for guide star
acquisition, dead cells, bad cells (\eg due to charge transfer
efficiency problems or dark noise) and the removal of portions of the
image by the \psone\ funding agency to excise fast moving satellites.
We expect the overall fill-factor due to all these effects to result
in $f\sim 0.88$.  As described above, for solar system discoveries we
require 6 detections on 3 nights within a lunation.  Thus, the impact
of the fill-factor on the detection efficiency ($\epsilon$) for fast
moving objects like the impactor population is simply $\epsilon =
f^{6} \sim 0.46$ if the inactive area is uncorrelated on the focal
plane or $\epsilon = f^{3} \sim 0.68$ if it is correlated.  That is,
if the first detection of an object appears on an inactive area then
its second detection is also likely to appear on an inactive area in
the `correlated' case and unlikely to do so in the `uncorrelated'
case.  Losing ${1 \over 3}$ to ${1 \over 2}$ of potential discoveries
due to the fill-factor is unfortunate but the loss is mitigated for
slow moving distant objects because they appear near opposition in
successive lunations.  \psone\ has two opportunities to discover the
object in each of two successive lunations so the efficiency for
finding these objects will be $\ga 90$\%.  The problem is worse for
objects like NEOs and impactors that spend fewer lunations above the
detection threshold.

To determine the Pan-STARRS (\psone) survey efficiency for detecting
impacting asteroids as a function of time and size we divided our
sample of $\sim$130,000 synthetic impactors into six independent sets
that were assigned different absolute magnitudes ($H$) as shown in
Table \ref{tab.ImpactorSFD}.  We can assign any $H$ to the impactors
because the Bottke NEO model assumed that the size distribution of the
NEOs was independent of the orbital elements. The six selected sizes
span almost two orders of magnitude in diameter. The number of objects
of each size was selected to provide a good number of derived objects
(after processing through MOPS) and yet not so many objects as to be
prohibitively expensive in processing time. Each of these six
independent sets of impactors were then run through a simulation of
the \psone+MOPS system.

Figure \ref{fig.PS1SurveyEfficiency} shows that the efficiency of the
\psone\ survey at detecting impactors increases as a function of time and
impactor diameter.  In just four years of \psone\ operations it could
identify $\sim$85\% of all 1~km diameter objects that will impact the
Earth in the next 100 years.  Interestingly, it has a $\sim$74\%
chance of identifying a 1~km diameter that would impact {\it during}
the 4 year time span of the survey.  At first glance this is
surprising since \psone\ can detect a 1~km diameter asteroid at opposition
at a geocentric distance of $\sim$2.67~AU.  The objects should be
visible long before impact in the large volume of space surveyed by
the \psone\ system.  We will explore the reasons for the ineffectiveness
of the survey at detecting these large hazardous objects below.

The impactor discovery efficiency increases nearly linearly with time
for objects of 200~m diameter to $\sim$40\% efficiency in just four
years.  While this behavior cannot increase indefinitely we estimate
that after 12 years the efficiency may be $\sim$80\%.  This is an
interesting value because the Pan-STARRS PS4 system will have
4$\times$ the collecting area\footnote{Since the PS4 system employs 4
separate cameras viewing the same field the PS4 image stacks will have
$\sim$100\% fill-factor.} of the \psone\ system modelled here and thus
the efficiency curve for 200~m diameter objects for \psone\
corresponds to the 100~m diameter efficiency curve for PS4.  Thus,
during its anticipated 10 year survey mission PS4 alone may reach
nearly 90\% completion for objects $\ge$140~m diameter or larger that
will impact in the next 100 years.

\begin{table}[h]
\small
\begin{center}
\caption{Size frequency distribution of synthetic impactors used in
the \psone\ survey simulation.  The conversion between diameter and
absolute magnitude assumes an albedo of $p=0.14$ for NEOs
\citep{Stuart04}.}  \centering
\label{tab.ImpactorSFD}
\begin{tabular}{ccc}
\\
\hline \hline
diameter & absolute  & number \\
(meters) & magnitude &  \\
\hline
1000 & 17.75 & 1193  \\
500  & 19.25 & 3816 \\
200  & 21.25 & 4804 \\
\hline
100 & 22.75 &  10001 \\
50  & 24.25 &  25001 \\
20  & 26.25 &  85000 \\
\hline
Total & & 129815 \\
\hline
\hline
\end{tabular}
\end{center}
\end{table}
The {\it impact warning time} was defined by \citet{Chodas08} as the
time between impact and when the probability of an impact on the Earth
is calculated to be more than 50\%.  However, we believe that the
experience and response to the discovery of (99942) Apophis shows that
the {\it impact awareness time} for an object is considerably longer
than the impact warning time.  Once an object is discovered that has a
non-zero probability of impact it is observed obsessively at every
opportunity and impact calculations are regularly refined to monitor
the impact probability.  We define the impact awareness time as the
period between the identification of an impactor as a PHO
(MOID~$<0.05$~AU) and its time of impact.  As discussed above, almost
all the impactors in this study have MOID~$\ll 0.05$~AU long before
impact.

Fig.~\ref{fig.derivedMOIDatDiscovery} (top) shows the MOID for derived
objects at discovery (when at least 3 nights of observations have been
obtained in the course of a single lunation). Almost all the objects
will immediately be flagged as PHOs and therefore start the impact
awareness time clock.  The error on the MOIDs, the difference between
the MOID for the synthetic object and the MOID for the derived object,
are small after just a single lunation with typical
$\Delta$MOID$\sim$MOID$ \sim 10^{-3}$~AU.  As more observations are
acquired for these objects the impact probability should monotonically
increase to 100\%.

Figure~\ref{fig.ImpactAwarenessTimes} shows the impact awareness time
as a function of the impactor size.  Since the synthetic impactors
were designed to hit the Earth at random times over the next 100
years, and large objects can be detected long before impact, it is no
surprise that the impact awareness time for large discovered synthetic
objects is evenly distributed over 100 years
(Figure~\ref{fig.ImpactAwarenessTimes} top).  The apparent decrease in
the fraction of 1000~m objects with larger impact awareness times is
an artifact of statistics - fitting a line to the distribution shows
that it is consistent with being flat at the 1-$\sigma$
level. Surprisingly, the impact awareness times for small discovered
synthetic objects is also evenly distributed over 100 years
(Figure~\ref{fig.ImpactAwarenessTimes} bottom).  This is because when
\psone\ discovers an object and observes it on $\ge$3 nights over an
$\sim 10$~day period the orbit is good enough to accurately determine
the MOID and determine the impact awareness time.  Of course, it may
be difficult to obtain followup observations of the smallest objects
to refine the orbit and the impact probability calculation.

It is important to keep in mind that
Fig.~\ref{fig.ImpactAwarenessTimes} does not include the large spike
at zero warning time due to the {\it undiscovered objects}.  For
example, Figure~\ref{fig.PS1SurveyEfficiency} shows that $\sim$60\% of
200~m diameter objects remain undiscovered at the end of the four
years of \psone\ surveying.  Thus, the most probable awareness
(warning) time for the smaller objects is zero.  But when they are
discovered before the apparition in which they impact the warning time
can be many decades. As discussed above, to be 90\% effective at
eliminating the risk of an unanticipated impact in the next 100 years
for objects $>$140~m diameter requires a \psone-like system to survey
much longer or a more powerful system like PS4 \citep{Kaiser05} or the
system being developed by the LSSTC \citep{Ivezic07}.

While it is interesting to compute the efficiency with which a next
generation survey such as \psone\ can find impacting asteroids the
reality of the situation is that it is unlikely that any large objects
are on a collision course with the Earth in the next 100 years.  The
expected number of detected impactors, $N(D)$, is simply the product
of the efficiency, $\epsilon(D)$, for detecting impactors of diameter
$D$ and the size-frequency distribution of the population, $n(D)$.
The efficiency was already shown in Fig.~\ref{fig.PS1SurveyEfficiency}
for $D>20$~m.  We use \citet{Brown02}'s determination of the annual
cumulative number of objects striking the Earth's atmosphere:
$N(>D)=37 (D/meters)^{-2.7}$ which implies that objects in the size
range of 2008~TC$_3$ strike the Earth about twice per year.
Fig.~\ref{fig.ImpactorDetections} shows that unless the Earth is
extremely unlucky \psone\ will not detect a large ($D>20$~m) impacting
asteroid.  A best case linear extrapolation of $\epsilon(D)$ to
$D<20$~m using just the 20~m and 50~m diameter points suggests that
the likelihood of obtaining 3 nights of detections in the discovery of
even smaller impactors is extremely unlikely.  The efficiency is
decreasing faster with smaller diameter than the number of objects is
increasing.

While it is unlikely that \psone\ will single-handedly obtain enough
detections to determine a pre-impact orbit, the possibility of
detecting a small asteroid prior to impact (such as the impactor 2008
TC$_3$), or precovering the asteroid after impact using an orbit
derived from the bolide trajectory, is also interesting to
meteoriticists.  With sufficient advance notice it would be possible
to organize a ground-based team to observe the asteroid's atmospheric
entry, train space-based platforms on the impact location, and prepare
for recovery of meteorites.  Post-impact precovery of the object could
be useful to help determine the pre-impact size of the object and to
verify the pre-impact orbit determination for the bolides
\citep[\eg][]{Abe08}.

Simulating the automated detection by PS1+MOPS of smaller impactors is
difficult. At the current time MOPS operates by linking together
tracklets on 3 nights taken over the course of $\sim10$~days.  With
typical impact velocities of $\sim15$~km/s the best case scenario
requires that \psone+MOPS first detect the object at a distance of
$\sim13\times10^6$~km (approximately 10 days before impact).  At
\psone's assumed limiting magnitude of $V=22.7$ this requires that the
object be $>~3$m in diameter.  Even if this situation were to unfold,
variations in the object's apparent position and velocity on the sky
due to the topocentric motion of the observatory would likely render
the object difficult to link within the MOPS (though we have not yet
studied this scenario in detail).  Instead, we consider the
possibility that \psone\ will detect an impacting asteroid prior to
impact but on too few nights to determine a pre-impact orbit.

It is unlikely that the smallest objects ($<20$~m diameter) will be
discovered, and have good enough derived orbits to guarantee impact,
in any apparition except for the one in which they strike the Earth.
Thus, we determined the sky-plane distribution and rate of motion 1, 5
and 10 days before impact for all the objects in
Table~\ref{tab.ImpactorSFD}.

Once again, since the Bottke NEO model's orbit distribution is
independent of size, and assuming that we can extend the orbit
distribution of the large objects down to 1~m diameter (we have
already discussed that while there is reason to believe that the
Yarkovsky effect will modify the orbit distribution there is as of yet
no debiased orbit distribution for the small impactors), we can assign
any size or $H$ to the objects and quickly determine the ensemble's
apparent magnitude distribution.  The efficiency for detecting these
objects is then simply the ratio between the number that meet all our
search criteria (within the surveying region, $V<22.7$, assuming that
surveying takes place on only 75\% of nights due to weather, and with
a rate of motion within the detectability range) and the total number.
Fig.~\ref{fig.BolideDetectionEficiency} shows that the maximum
efficiency for the smallest objects is about 0.6\%.

The distributions of the rates of motion of the meteoroids before
impact are shown in Fig.~\ref{fig.BolideRateOfMotion}.  Typical rates
of motion are much slower on the final approach trajectory than on
fly-by apparitions.  The detectable rates of motion are limited to $0
\le \omega \le 12$~deg/day where the upper limit is enforced by the
\ps\ funding agency.  The figure shows clearly that the restricted
rate of motion does not have a strong effect on detection efficiency
when the impactors are on their approach to the Earth.

In the discussion above and elsewhere in this work we have ignored the
effects of trailing loss --- when objects move during the exposure time
they leave `trails' on the image rather than point sources and the
trails have, in the past, been more difficult to identify than point
sources.  While the \psone\ point source detection limit is expected
to be $V \sim 22.7$ the efficiency of our faint trail detection
algorithm is not yet known.  It is thought that it should be efficient
to per-pixel $S/N \ll 1$ implying that the detection efficiency could
be quite good even for intrinsically faint objects.  Since we have not
yet measured the trailing losses we ignore them.  This means that the
small impactor detection efficiencies reported here and in
Fig.~\ref{fig.BolideDetectionEficiency} are upper limits.

Figure~\ref{fig.ImpactorDetections} shows the expected number of
detections of small impactors (bolides) within 10 days of impact
during the 4 year \psone\ survey as a function of diameter using the
\citet{Brown02} size-frequency distribution (SFD).  The cumulative
probability of \psone\ detecting a bolide in the 1-10~m size range is
$25\pm15$\% (the error is determined from the SFD only). In other
words, \psone\ has $\sim 25$\% probability of detecting another
2008~TC$_3$-like object during its four year survey mission.

Considering that we have argued that \psone\ is the first of the next
generation sky-surveys and that this powerful system only has a $\sim
25$\% probability of detecting another 2008~TC$_3$-like object in four
surveying years, how could a `last-generation' survey like CSS have
discovered 2008~TC$_3$? In addition to the impactor, the CSS has also
discovered several other small asteroids making very close approaches
to the Earth (\eg 2008 UA$_{202}$).  The CSS's telescope system has
(Ed Beshore, personal communication) a smaller etendue (1.5~m $\times$ 1
deg$^2$) than the \psone\ system (1.8~m $\times$ 7 deg$^2$) at a
significantly worse site (CSS FWHM $\sim 2\arcsec$ vs $\sim 1\arcsec$
for \psone; worse weather conditions; brighter sky), and twice the
camera readout time ($\sim 12$~s vs. $\sim 6$s). Thus, we
conservatively estimate that the \psone\ system will be $\sim
10\times$ more efficient than the CSS.  The CSS survey strategy
sacrifices limiting magnitude to cover as much sky as possible (T.\
Spahr, personal communication) --- an old recipe for finding more NEOs
\citep{Bowell94}. The result is that the CSS 1.5~m telescope covers
about 3600 deg$^2$ per lunation to $V\sim21$ mag while our \psone\
{\it simulation} covers $\sim$3800 deg$^2$ to a limiting magnitude of
$V\sim22.7$.  Based on these arguments our simulations suggest that
the CSS made a low-probability discovery of 2008~TC$_3$ --- they were
lucky.  Another alternative is that the SFD of the bolides from
\citet{Brown02} is in error --- we consider this to be unlikely because
of other corroborating research \citep[\eg][]{Ivanov06}.

Like the CSS discovery of 2008~TC$_3$, the problem is that \psone\
will detect the object on only a single night resulting in an orbit
with large uncertainties. \psone\ can report single night detection
pairs to the Minor Planet Center (MPC) but will then depend on rapid
announcement by the MPC of likely impactors and prompt followup by
other observatories around the world.  

We have considered the possibility that detections in the impactor's
tracklet could provide information to flag it as an imminent impactor
based on rapid changes in the detections' trail length or orientation
due to the effects of topocentric parallax.  While the 15-30~min time
difference between the detections in the tracklets probably does not
provide enough leverage to make this method viable for Pan-STARRS it
could be employed in the future by a dedicated impactor survey.

Since ground-based all-sky bolide impact monitoring networks
\citep[\eg][]{Oberst98,Weryk08,Bland08} cover an extremely small
fraction of the Earth's surface ($\sim$1\%; P. Brown, personal
communication) it is exceedingly unlikely that an object imaged by
\psone\ will also be detected by a ground-based sensor.  On the other
hand, the detection of bolides from wide-coverage space-based systems
\citep[\eg][]{Koschny04} is possible and may provide an opportunity to
search through \psone\ detections for pre-impact observations.

Figure~\ref{fig.bolide_skyplane} shows the apparent sky-plane
distribution for 3~m diameter bolides 1 day before impact.  These
small objects (\eg~1-20~m) are visible for only a short time before
impact and their apparent brightness is strongly affected by the phase
angle so that they can only be detected near opposition. The location
of 2008~TC$_3$ 3 days before impact is clearly within the \psone\
opposition field.  \psone\ would have detected 2008~TC$_3$ since it
was brighter than V=22.7~mag (assuming $H=30.4$ and $G=0.15$) for 4--5
days before impact (with $V=18.9$~mag at discovery ~1 day before the
impact) and the sky-plane motion was below the \psone\ trailing
cut-off of 12$\deg$/day.

Although the small impactors may be located over the entire sky there
are two prominent clumps in Fig.~\ref{fig.bolide_skyplane} --- towards
the Sun and around the opposition point (the anti-solar
direction). This result is consistent with the helion and antihelion
sources \citep{Poole97} of sporadic background bolides that are the
result of the velocity vector summation of the Earth and the sporadic
meteors. \citet{Jones93} had earlier suggested that these two sources
of sporadic meteors were most consistent with a population of small
meteoroids derived from a Jupiter Family Comet (JFC) parent
population. We note that their study did not correct for strong
selection effects in the radar detection of the meteors that would
favor the high velocity cometary sub-component of the helion and
anti-helion sources.  Considering that the Bottke NEO model employed
here is dominated by asteroids but still produces these two sources it
suggests that at least some component of the sporadic meteors in the
helion and anti-helion regions may be of asteroidal origin.

We believe that our results encourage optimism that the suite of next
generation ground-based sky surveys will be able to eliminate
$\ge$90\% of the risk of an unanticipated impact of an object $>140$~m
diameter. Still, it is interesting to identify a survey strategy that
could reduce the risk even further.  Figure~\ref{fig.unfound_1km}
shows that the undiscovered large (1~km) impactors rarely appear in
the \psone\ survey regions and are strongly clustered in the direction
towards the Sun. This is not because the objects are on orbits
interior to the Earth's orbit \citep[\eg][]{Zavodny08} but is due to
their orbital period being close to one year.
The synodic period with respect to Earth $P_{syn} =
|P_{Earth}^{-1}-P_{asteroid}^{-1}|^{-1}$ is thus long. The four year
\psone\ survey has a low efficiency for detecting objects with long
synodic periods and the solution is to survey over a longer time
period or to survey the sky at smaller solar elongations.  The suite
of next generation surveys (\eg, \ps, LSSTC) will survey the sky for
about a decade and will increase the completion statistics at all
impactor sizes. Surveying closer to the Sun is realistic only from
future space-based platforms
\citep[\eg][]{Tedesco00,Jedicke03,Hildebrand07,Mottola08}.

Figure~\ref{fig.1kmOrbitDistn} shows the orbital period for 1~km diameter
impactors that strike the Earth {\it during} the 4 year survey
mission.  The orbital period of the undetected impactors peaks at 1
year as explained above.  At first glance it may be surprising that a
powerful survey system could miss these large impactors given the
large distance at which 1~km diameter objects can be detected.  The
explanation is that all the undetected large impactors strike the
Earth in the first 2 years; before the survey had sufficient time to
explore the entire volume of the solar system to the distance at which
1~km diameter objects are detectable.

\section{Conclusions/Discussion}\label{concl}

We have used a synthetic population of Earth-impacting asteroids to
determine their sky-plane distribution and detectability with one of
the next generation all-sky surveys, \psone.

We find that
\begin{itemize}
\item while the steady state sky plane distribution of the impactors
  is concentrated towards small solar elongation as shown by
  \citet{Chesley04} their sky-plane distribution exhibits interesting
  behavior in the couple years leading up to impact.  This behavior
  could be exploited to identify these objects well in advance of
  impact.
\item the requirement that a PHO have a MOID$<$0.05~AU identifies 98\%
  of all impactors even 100 years in advance of impact.  The remaining
  2\% have much larger MOID and the error on the MOID will not
  identify these objects as potential impactors since they become PHOs
  only after encounter with another solar system object, usually
  Jupiter.  However, numerical exploration of their orbit evolution as
  routinely performed by impact monitoring sites will generally reveal
  these hazardous objects long before impact.
\item the MOIDs determined for objects with detections obtained during
  just a single lunation are very good
\item the impactor model that we have developed is probably a good
  proxy for the unbiased orbit distribution of the large bolide
  population based on 1) the good agreement between the impactor orbit
  distribution and the orbit elements for impacting asteroid
  2008~TC$_3$ and 2) its sky-plane location at the time of discovery.
\item the next generation of all-sky surveys will identify a large
  fraction ($\ga 90$\%) of impactors $>$140~m diameter in a
  decade-long survey designed to find them.
\item the impact awareness time before impact can be many decades for
  impactors of any size as long as they are discovered before impact.
  The most likely impact awareness time for smaller impactors is zero.
\item the next generation surveys will probably image a small impactor
  before impact but it will likely be observed over too short a time
  span (\eg one night) to allow either an accurate pre-impact orbit to
  be computed or its identification as an imminent impactor.
\item a search for pre-impact detections of a bolide is possible
\item the most difficult objects to discover are those with long synodic
  periods relative to the Earth.
\item next generation surveys like \psone\ will be efficient at
  detecting large impactors when their impact time is more than about
  2 years after the start of the survey.
\end{itemize}


\clearpage

\section*{Acknowledgements}

This work was performed in collaboration with Spacewatch, LSSTC, CMU's
AUTON lab, and the AstDyS team (Andrea Milani, Giovanni Gronchi and
Zoran Knezevic).  Steve Chesley's work was conducted at the Jet
Propulsion Laboratory, California Institute of Technology, under a
contract with the National Aeronautics and Space Administration.  The
design and construction of the Panoramic Survey Telescope and Rapid
Response System by the University of Hawaii Institute for Astronomy is
funded by the United States Air Force Research Laboratory (AFRL,
Albuquerque, NM) through grant number F29601-02-1-0268.  MOPS is also
supported by a grant (NNX07AL28G) to Robert Jedicke from the NASA NEOO
program.  Andrea Milani, Giovanni Gronchi and Zoran Knezevic of the
AstDyS group provided critical orbit determination software to the
MOPS team.  The MOPS is currently being developed in association with
the Large Synoptic Survey Telescope Corporation (LSSTC). The LSSTC's
research and development effort is funded in part by the National
Science Foundation under Scientific Program Order No. 9 (AST-0551161)
through Cooperative Agreement AST-0132798. Additional funding to the
LSSTC comes from private donations, in-kind support at Department of
Energy laboratories and other LSSTC Institutional Members.  Veres was
supported by the National Scholarship Programme of the Slovak
Republic, the European Social Fund, a Grant from Comenius University
(No. UK/399/2008) and VEGA Grant No. 1/3067/06.


\clearpage

\clearpage

\clearpage

\clearpage

\clearpage

\clearpage

\clearpage

\clearpage
\addtocounter{figure}{-1}
\begin{figure}[htp]
\centering
\includegraphics[width=0.9\textwidth]{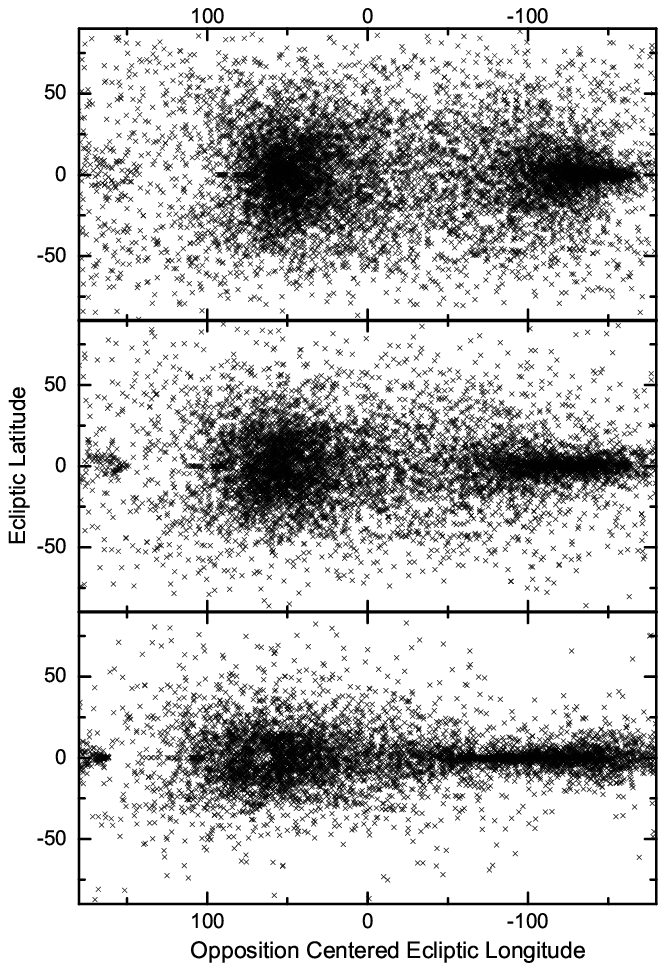}

\caption{b) The opposition centered ecliptic sky plane distribution of
Earth-impacting asteroids with $V \le 22.7$ ({\it top}) 90 days, ({\it
middle}) 120 days, and ({\it bottom}) 150 days before impact.}

\label{fig.skyplaneTimeEvolution-b}
\end{figure}

\clearpage
\addtocounter{figure}{-1}
\begin{figure}[htp]
\centering
\includegraphics[width=0.9\textwidth]{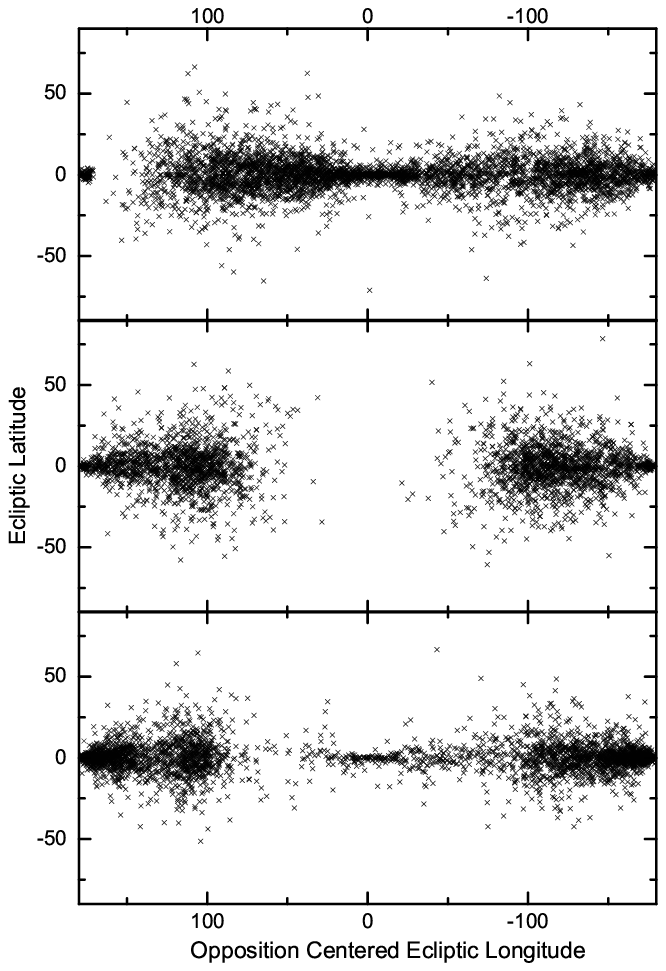}

\caption{c) The opposition centered ecliptic sky plane distribution of
Earth-impacting asteroids with $V \le 22.7$ ({\it top}) 180 days, ({\it
middle}) 1 year, and ({\it bottom}) 1.5 years before impact.}

\label{fig.skyplaneTimeEvolution-c}
\end{figure}

\clearpage
\addtocounter{figure}{-1}
\begin{figure}[htp]
\centering
\includegraphics[width=0.9\textwidth]{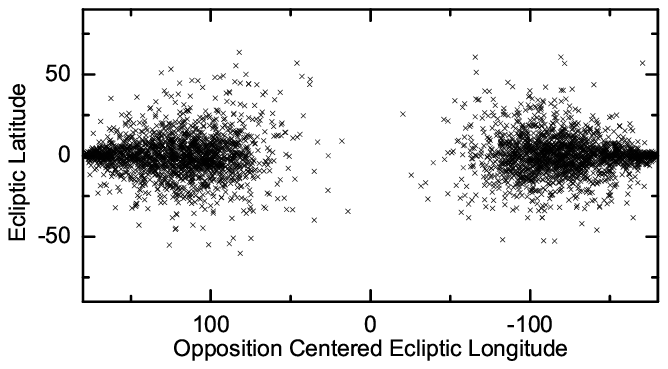}

\caption{d) The opposition centered ecliptic sky plane distribution of
Earth-impacting asteroids with $V \le 22.7$ 2 years before impact.}

\label{fig.skyplaneTimeEvolution-d}
\end{figure}

\clearpage
\begin{figure}[htp]
\centering
\includegraphics[height=120mm, width=180mm]{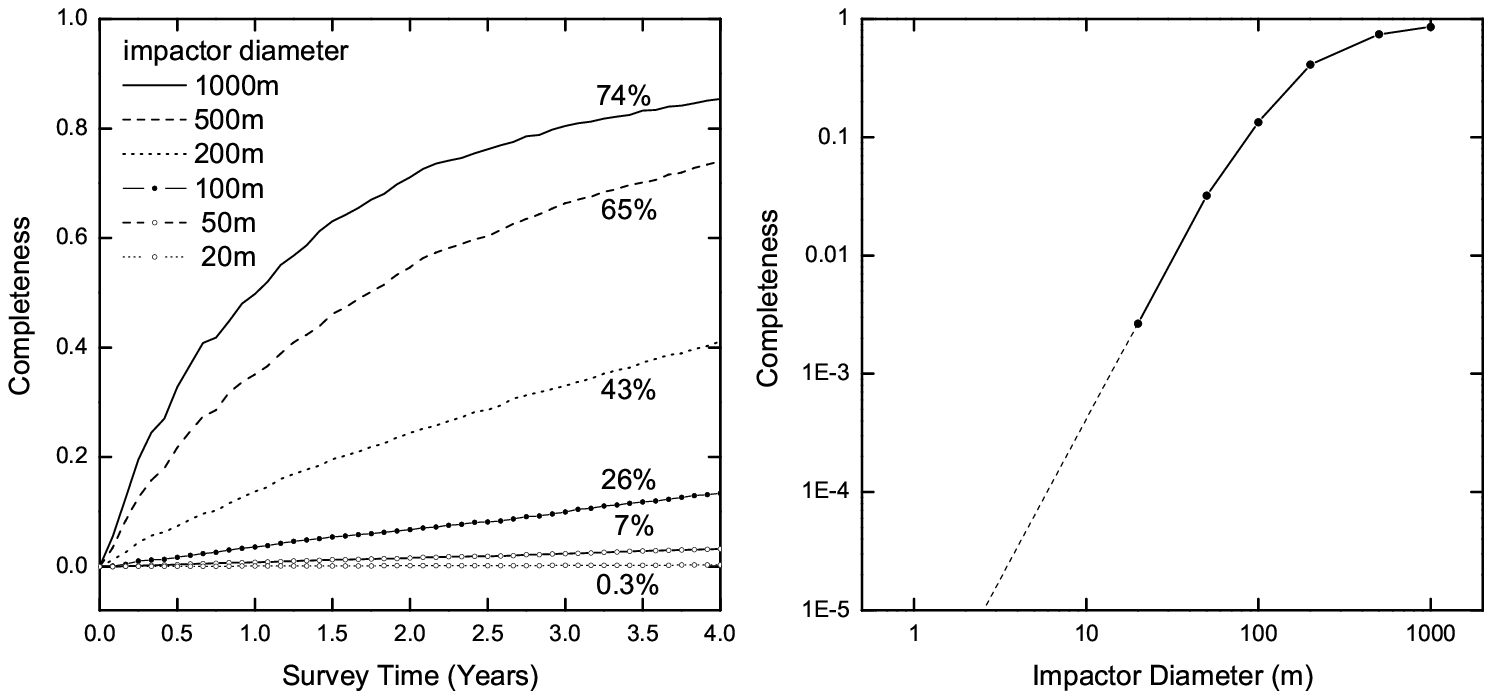}

\caption{({\it left}) \psone\ survey completeness (efficiency) as a
function of time for different size impactors assuming an albedo of
$0.14$.  The percentages on the right side of the figure adjacent to
the efficiency curves are the efficiency for detecting an impactor
that strikes the Earth {\it during} the 4 year survey. ({\it right})
The efficiency for discovering objects impacting during the 4 year
survey as a function of impactor diameter.  A linear extrapolation to
$D<20$~m using the two leftmost data points is shown with a dashed
line.}

\label{fig.PS1SurveyEfficiency}
\end{figure}

\clearpage
\begin{figure}[htp]
\centering
\includegraphics[width=1.1\textwidth]{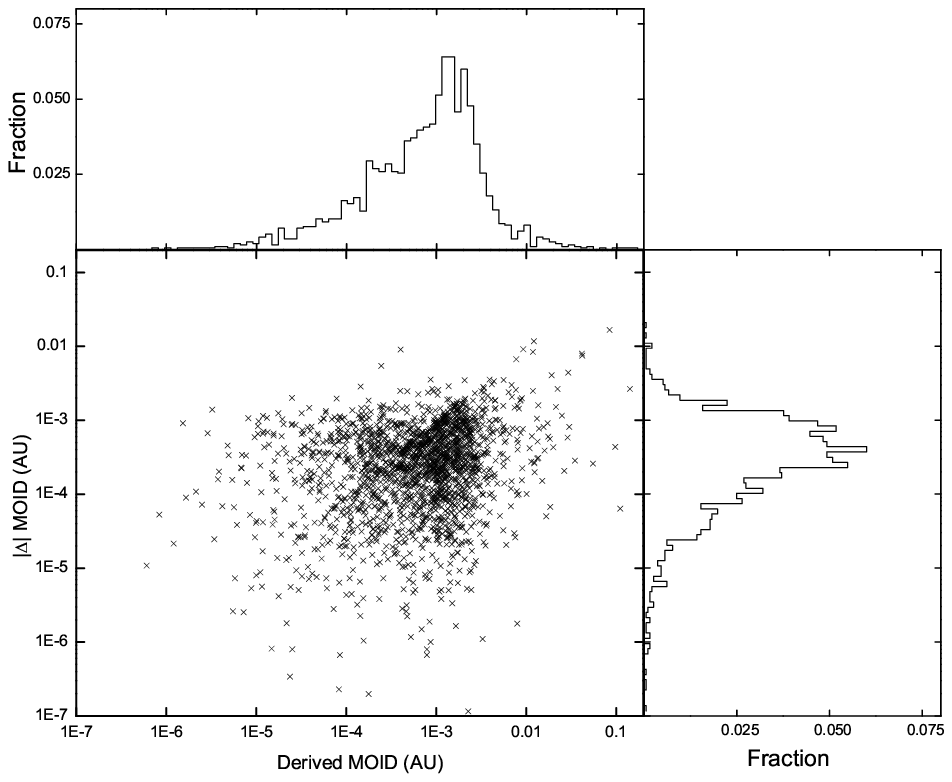}

\caption{({\it lower-left}) The distribution of the error in the MOID
($|$MOID$_{synthetic}-$MOID$_{derived}|$) vs. the derived MOID for
impactors detected in the simulation. ({\it right}) Distribution of
the error in the MOID. ({\it top}) Distribution of the derived MOID.}
\label{fig.derivedMOIDatDiscovery}
\end{figure}

\clearpage
\begin{figure}[htp]
\centering
\includegraphics[height=100mm, width=120mm]{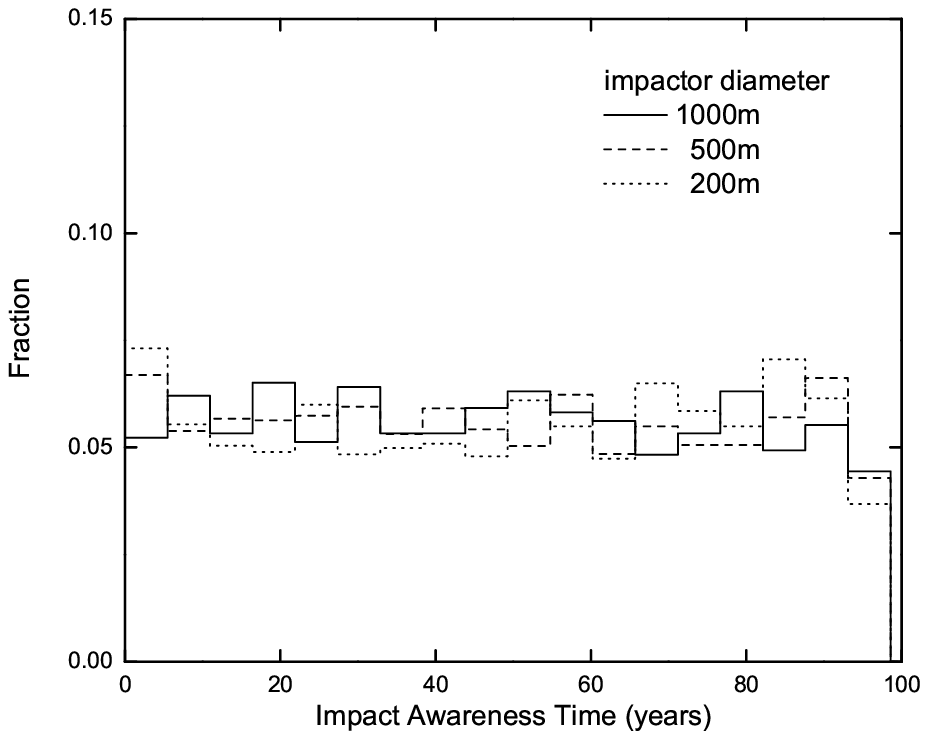}
\includegraphics[height=100mm, width=120mm]{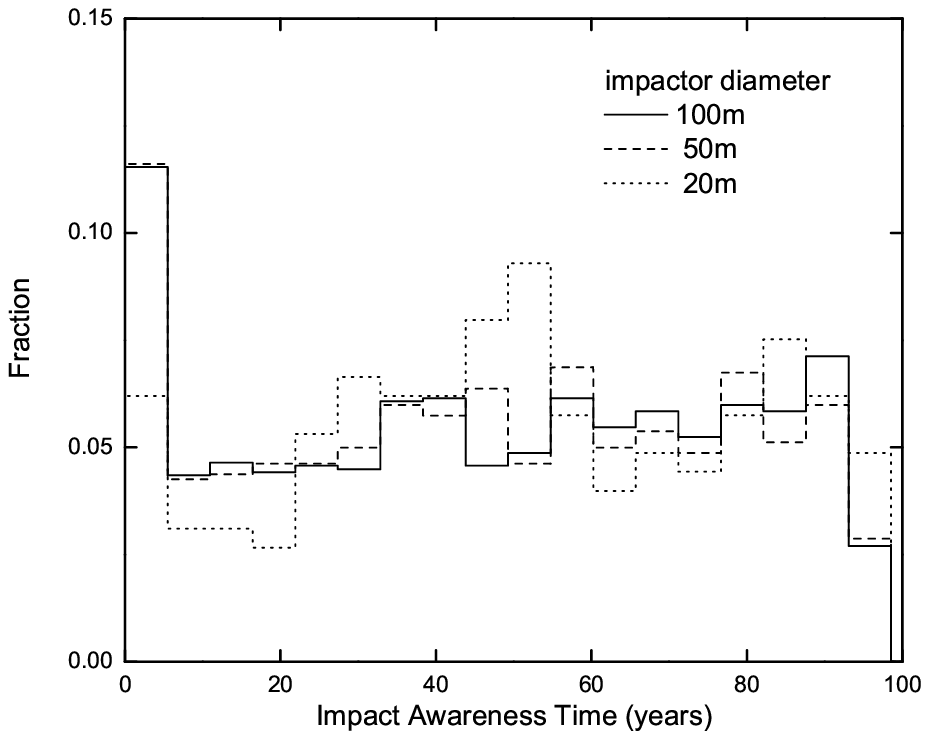}

\caption{Distribution of impact awareness times (defined in the text)
for discovered synthetic impactors of different sizes.  Note that the
awareness time is zero for undiscovered impactors and they are not
included in this figure.}

\label{fig.ImpactAwarenessTimes}

\end{figure}

\clearpage
\begin{figure}[htp]
\centering
\includegraphics[height=120mm, width=160mm]{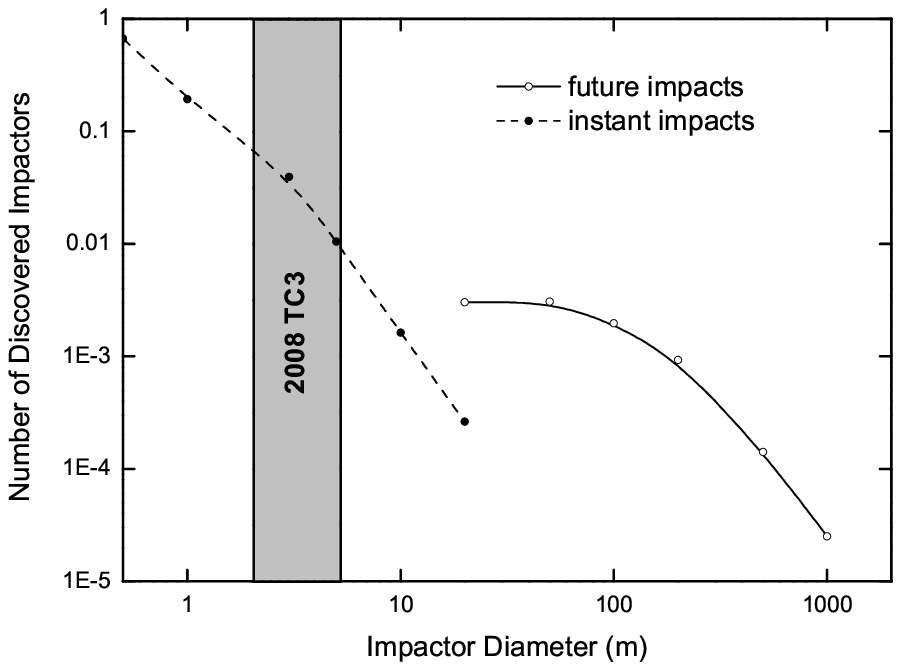}

\caption{({\it solid}) Expected number of detections of large ($>20$~m
diameter) impactors that will strike the Earth in the next 100 years
that are observed 3 nights in a single lunation during a four year
\psone\ survey and ({\it dashed}) the number of small (1-20~m
diameter) impactors detected by the \psone\ system in the lunation
prior to impact. The probable size range of the impacting asteroid
2008~TC$_3$ is indicated by the shaded region.}

\label{fig.ImpactorDetections}

\end{figure}

\clearpage
\begin{figure}[htp]
\centering
\includegraphics[height=120mm, width=150mm]{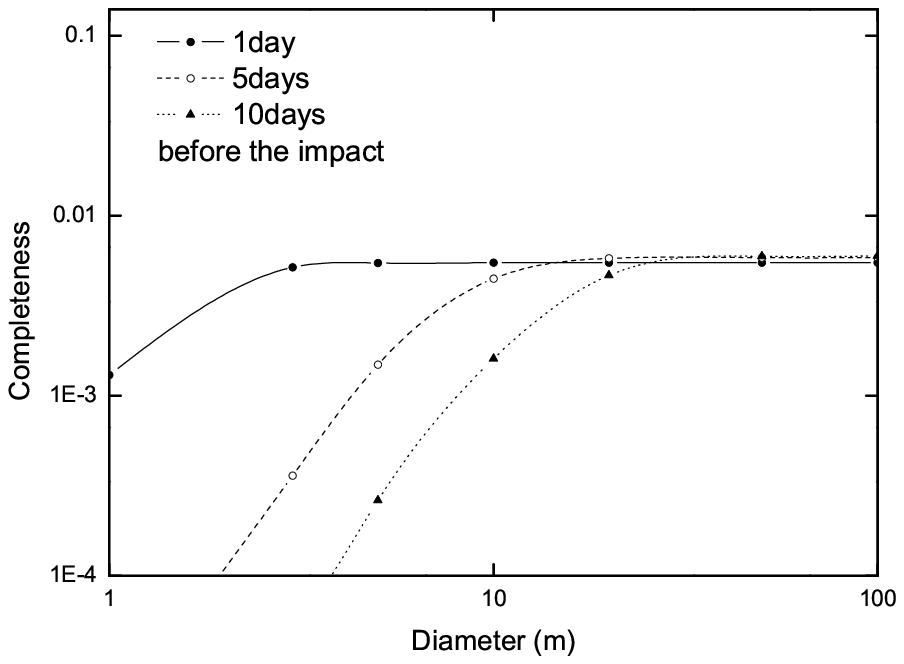}

\caption{Estimated four year \psone\ survey detection efficiency 1, 5, and
10 days before impact for small impactors on their final approach to
the Earth.}

\label{fig.BolideDetectionEficiency}

\end{figure}

\clearpage
\begin{figure}[htp]
\centering
\includegraphics[width=0.9\textwidth]{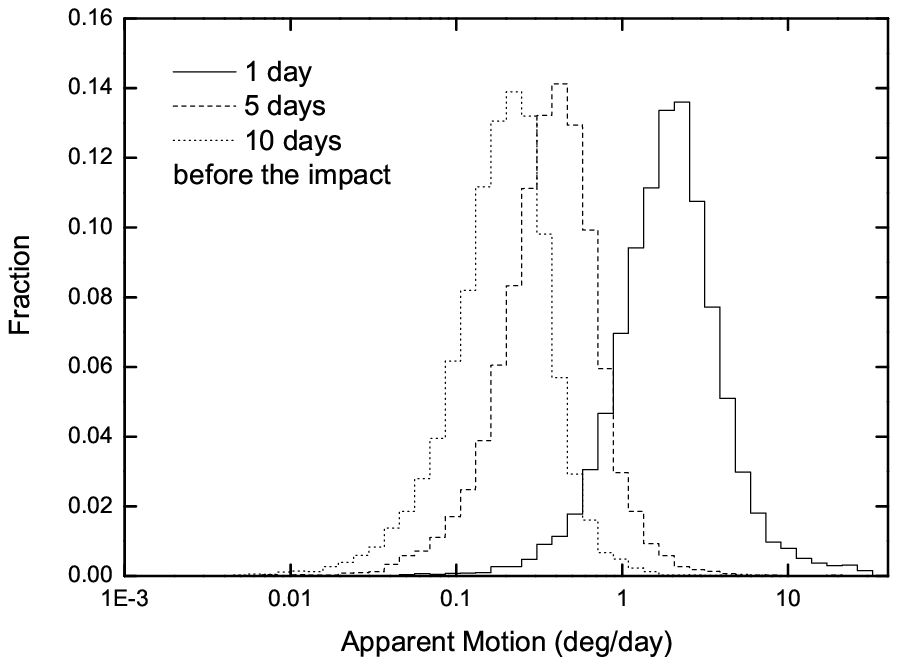}

\caption{Distributions of the apparent rate of motion for small
impactors 1, 5, and 10 days before impact.  There is no restriction on
apparent magnitude.  The upper limit on the detectable rate of motion
imposed by the \ps\ funding agency is 12 deg/day.}

\label{fig.BolideRateOfMotion}

\end{figure}

\clearpage
\begin{figure}[htp]
\centering
\includegraphics[width=0.9\textwidth]{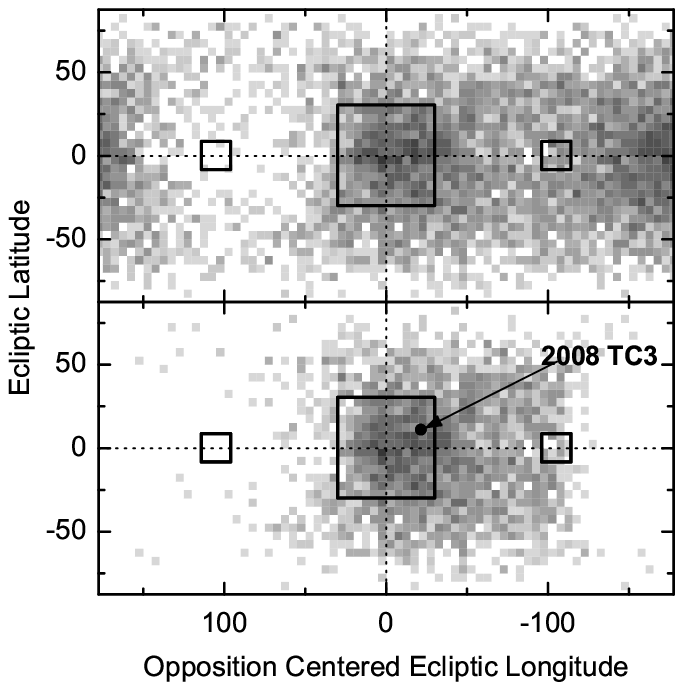}

\caption{({\it top}) Sky-plane probability distribution for 3~m
diameter impactors (bolides) 1 day before impact with no restrictions
on brightness and apparent motion.  ({\it bottom}) Same as above but
with $V < 22.7$ and apparent motion $<12$~deg/day.  Approximate
\psone\ search regions are shown as solid rectangles (the opposition
region is in the center and the sweet spots are on the left and
right).  The position of 2008~TC$_3$ at discovery by the Catalina Sky
Survey is highlighted.}

\label{fig.bolide_skyplane}

\end{figure}

\clearpage
\begin{figure}[htp]
\centering
\includegraphics[width=0.9\textwidth]{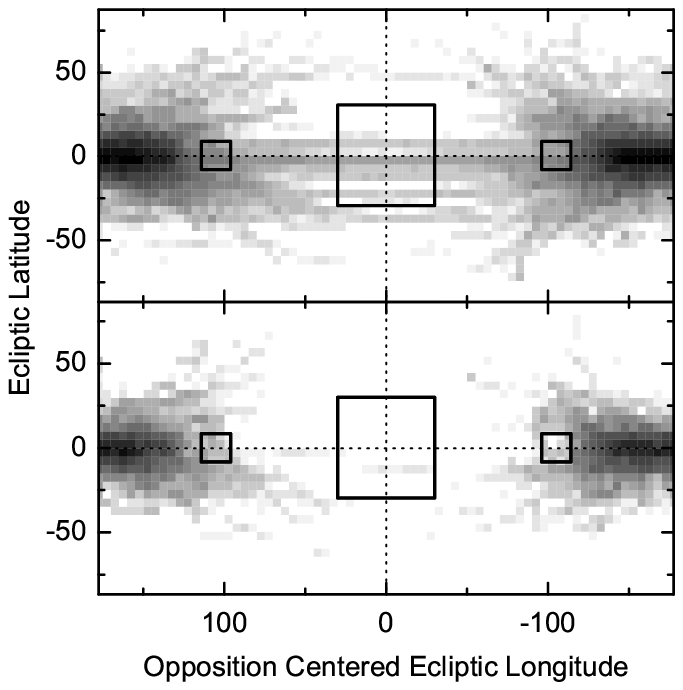}

\caption{({\it top}) Sky-plane position probability distribution for
1~km diameter impactors that were {\it not} discovered during the
simulated 4 year \psone\ survey mission with no restriction on the
apparent magnitude.  Darker regions represent where the objects are
most likely to be found.  ({\it bottom}) Same as above but imposing a
\psone\ limiting magnitude cutoff at $V=22.7$. Approximate \psone\
search regions are shown as solid rectangles.}

\label{fig.unfound_1km}
\end{figure}

\clearpage
\begin{figure}[htp]
\centering
\includegraphics[width=1.0\textwidth]{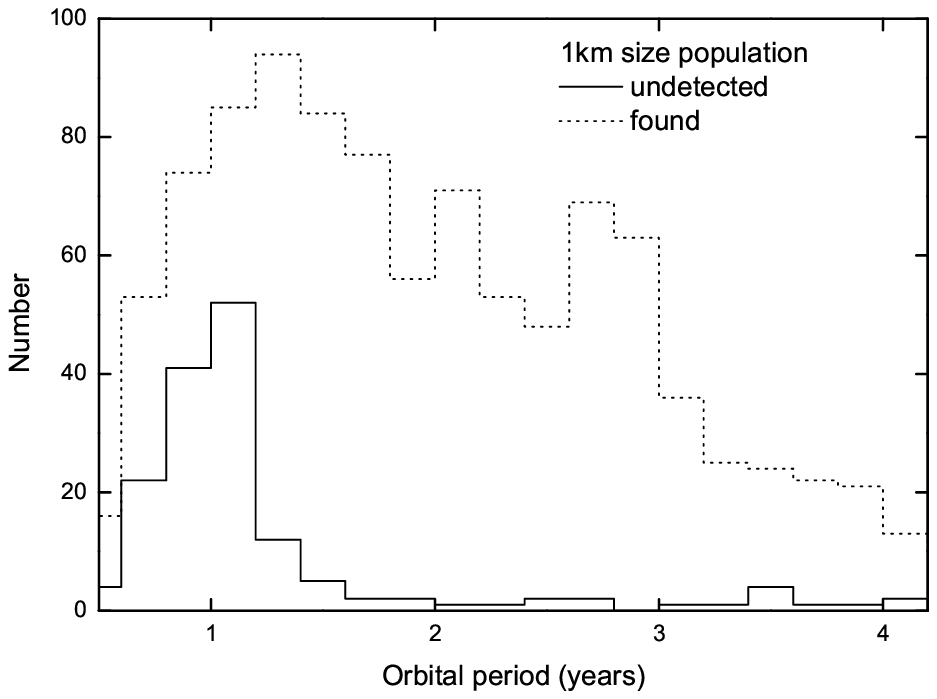}

\caption{Orbital period distribution for found ({\it dotted}) and
undetected ({\it solid}) 1~km diameter impactors during the 4 year
\psone\ survey mission.}

\label{fig.1kmOrbitDistn}
\end{figure}

\end{document}